\documentclass[a4j,11pt]{article}
\usepackage[dvipdfmx]{graphicx}
\usepackage{epstopdf, epsfig}
\usepackage{bm}
\usepackage{amsmath}
\usepackage{amssymb}
\usepackage{mathrsfs}
\usepackage{amsmath}
\usepackage{here}
\usepackage{natbib}

\usepackage{color}

\newtheorem{lemma}{Lemma}

\newtheorem{theorem}{Theorem}
\newtheorem{proposition}{Proposition}

\newcommand{\mathsfbi}[1]{\mathrm{\textrm{#1}}}
\def \bcdot{\cdot}
\def \bnabla{\nabla}

\newcommand{\qed}{\nobreak \ifvmode \relax \else
      \ifdim\lastskip<1.5em \hskip-\lastskip
      \hskip1.5em plus0em minus0.5em \fi \nobreak
      \vrule height0.75em width0.5em depth0.25em\fi}
\newcommand{\QED}{\begin{flushright} ~\qed \end{flushright}}

\title{Partition of enstrophy between zonal and turbulent components}

\author{Hiroto Aibara and Zensho Yoshida\\
~\\
Graduate School of Frontier Sciences, The University of Tokyo, \\
Kashiwa, Chiba 277-8561, Japan}

\begin{document}

\maketitle

\begin{abstract}
The partition of enstrophy between zonal (ordered) and wavy (turbulent) components of vorticity  has been studied for the beta-plane model of  two-dimensional barotropic flow.
An analytic estimate of the minimum value for the zonal component has been derived.
The energy, angular momentum,  circulation, as well as the total enstrophy are invoked as constraints for the minimization of the zonal enstrophy.
The corresponding variational principle has an unusual mathematical structure (primarily because the target functional is not a coercive form), by which the constraints work out in an interesting way.  
A discrete set of \emph{zonal enstrophy levels} is generated by the energy constraint; 
each level is specified by an eigenvalue that represents the lamination period of zonal flow.
However, the value itself of the zonal enstrophy level is a function of only angular momentum and circulation,
being independent of the energy (and total enstrophy).
Instead, the energy works in selecting the ``level'' (eigenvalue) of the relaxed state.
The relaxation occurs by emitting small-scale wavy enstrophy, and continues as far as the nonlinear effect, scaled by the energy, can create wavy enstrophy.
Comparison with numerical simulations shows that the theory gives a proper estimate of the zonal enstrophy in the relaxed state.
\end{abstract}


\section{Introduction}
\label{sec:introduction}
The creation of \emph{zonal flow} in the planetary atmosphere is a spectacular example of the self-organization in physical systems\,\citep{Hasegawa}.
There is a strong analogy between the geostrophic turbulence and the electrostatic turbulence of magnetized plasma in the plane perpendicular to an ambient magnetic field;
this similarity manifests as the mathematical equivalence of the Charney equation of Rossby waves\,\citep{Charney} and the Hasegawa-Mima equation of plasma drift waves\,\citep{HasegawaMima}.
Because the generation of zonal flow (coherent structure) affects the turbulent transport in magnetized plasmas, how strong it can be is of great interest in the context of plasma confinement\,\citep{Diamond}.
The aim of this work is to estimate the minimum value of the ``zonal enstrophy'' and elucidate how it is determined;
the minimum zonal enstrophy indicates that the zonal flow must be stronger than its value.

The inverse-cascade model explains the essence of the self-organization process.
Because of the approximate two-dimensional geometry 
(due to the scale separation between the shallow vertical direction and wide horizontal directions), 
the vortex dynamics is free from the stretching effect.
Then, the energy of flow velocity tends to accumulate into large-scale vortices,
while the enstrophy (the norm of vorticity) cascades to small scales\,\citep{Kraichnan}.
On a rotating sphere, the gradient of Coriolis force yields the Rossby-wave term in the vortex dynamics equation,
which brings about latitude / longitude anisotropy, and the large-scale vorticity form zonal flow\,\citep{Charney}.
The nonlinear term driving the inverse cascade becomes comparable to the linear Rossby-wave term at the \emph{Rhines scale} which gives a crude estimate of the latitude size of the zonal flow\,\citep{Rhine}. 

While the inverse-cascade model illustrates the general tendency of nonlinear process, the underlying mechanism requires more detailed analysis. 
The modulational instability plays an essential role in exciting the energy transfer in the wave-number space\,\citep{Lorentz,Gill,CNNQ}.
In addition to the energy and enstrophy, another quadratic integral is known to be an adiabatic invariant (only changes by fourth-order of perturbations) in the zonal-flow domain of wave-number space, 
restricting the energy transfer there\,\citep{BalkNazarenkoZakharov,Balk,Balk2}. 
Various numerical simulations have been done to demonstrate the creation of zonal flow. 
Two different categories of models must be distinguished; one is the unforced, free decaying turbulence, and the other is the forced, quasi-stationary turbulence.
In the latter case, the interaction between the mean flow and the turbulence\,\citep{FarrellIannou,BakasIoannou,SrinivasanYoung} or inhomogeneous vorticity mixing\,\citep{DritschelMcIntyre,ScottDritschel} have been found as causal mechanisms of zonal flow generation.
 For these forced, quasi-stationary cases, one has to include some dissipation mechanism for large scale flows in order to remove the energy accumulating in the large scale regime by the inverse cascade.
The usual viscosity only works for short scale flows, so something like ``friction'' is added to the model
(however, which mechanism works in a realistic planetary system is still controversial).
For the free decaying case, early simulation results\,\citep{YodenYamada, VallisMaltrud, YIHY} 
demonstrated the self-organization of zonal flow, and found that the scale of zonal flow has similar scaling with Rhines' estimate. 
However, the quantitative comparison between the Rhines scale and the zonal flow scale was left unclear. 
On the other hand, in the forced turbulence case, more complex relation has been found, because of the influence of the dissipation mechanism for large scale flows; see \,\citep{Williams, DanilovGurarie, SukorianskyDikovskaya}.


In parallel with simulation studies, there have been theoretical attempts to nail down the ``target'' of the spontaneous process,
i.e. formulating a variational principle that reveals what the dynamics tends to reach.
This can be done by identifying the target functional to be minimized (or maximized) as well as the constraints that restrict admissible candidates.
A well-known example is the entropy maximization in the microcanonical ensemble
(the constraints are total particle number and total energy), which gives the Gibbs distribution. 
In the application to field theories, where we have to deal with infinite-dimension phase spaces,
we encounter the problem of ultraviolet catastrophe (which must be removed by appropriate quantization\,\citep{Ito-Yoshida}). 
Suspending such subtle problems, formal calculations have been made to obtain the thermal equilibrium distribution of flow fields.
In the context of the planetary atmosphere, the statistical equilibrium state in the two-dimensional incompressible Euler system was studied in \,\citep{miller, RobertSommeria}.
In \citep{TMHD}, the maximum entropy distribution over the ensemble constrained by total energy and circulation was compared with the large-scale vortex structures observed on Jupiter.
However, because of the essential non-equilibrium property of turbulence
(as the cascade model is based on ``dissipation'' in the Kolmogorov microscales,  one has to assume a ``driving force'' to maintain the (quasi) stationary state, or consider a transient process of free decay), the entropy may not be an effective tool to dictate the self-organization.

There is a different type of approach guided by the notion of \emph{selective dissipation}\,\citep{Hasegawa}.
We begin by making a list of conservation laws that apply in the \emph{ideal} (i.e. dissipation-less) model.
A finite dissipation breaks most of the conservation laws.
However, there may be differences in fragility among the constants of motion. 
Here, \emph{fragility} means the sensitivity to small scale perturbations (effective in the Kolmogorov microscale);
as the antonym, we say \emph{robust} if the constancy is unaffected by small scale perturbations.
We can estimate the fragility by counting the number of spacial derivations included in the ideal constants (see Appendix A).
If the most fragile one decreases (or increases) monotonically,
we choose it for the target functional, and the others for (approximate) constraints.
The minimization (or maximization) may parallel the \emph{relaxation process};
this is, of course, a very crude model of the complex dynamics, assuming that the chosen small-number constants only dictates the relaxation process. 

The \emph{Taylor state} of magneto-fluid\,\citep{JBTaylor,JBTaylor_review} is the prototype of such a model of self-organization,
which minimizes the magnetic field energy
($E = \frac{1}{2} \int |\boldsymbol{B}|^2\,\mathrm{d}^3x = \frac{1}{2} \int |\bnabla\times\boldsymbol{A}|^2\,\mathrm{d}^3x$, where $\boldsymbol{B}=\bnabla\times\boldsymbol{A}$ is the magnetic field) 
under the constraint on the magnetic helicity
($H=\frac{1}{2} \int \boldsymbol{A}\bcdot\boldsymbol{B}\,\mathrm{d}^3x$).
The reason why $E$ is more fragile than $H$ is because $E$ includes another differential operator curl in the integrand.
This model explains the \emph{relaxed states} of magnetized plasmas in various systems, ranging from laboratory experiments to astronomical objects.
Regarding the two dimensional turbulence in the planetary atmosphere, the minimization of the generalized enstrophy (see Proposition\,\ref{proposition:constants}) under the constraint on the energy has been studied to show that the solution of the minimization problem predicts a steady state with streamlines parallel to contours of the topography\,\citep{BH76}.
Although these two stories, i.e. the energy-helicity  relation in the magneto-fluid and the enstrophy-energy relation in the 2D-fluid appear to be parallel
(as \citet{Hasegawa} describes in the unified vision),
there is a fundamental difference when viewed from their Hamiltonian structures,
and the latter needs a careful interpretation.
 In both systems, the ideal constants (the helicity in magneto-fluids and the enstrophy in 2D-fluids) are Casimirs,
 by which the orbits are constrained on the level-sets of these constants\,\citep{Morrison}.
 In the magneto-fluid phase space, the orbits converge into the equilibrium point as the energy diminishes; the minimum energy (Hamiltonian) state, on each level-set of the helicity, gives an equilibrium point.
In the 2D-fluid system, however, the level set of the enstrophy is not embedded as a smooth submanifold in the topology of the energy norm (because the enstrophy is a fragile quantity, its level-set looks like a fractal set; see Appendix A).
Hence, we have to reverse the role of the Hamiltonian (energy) and the Casimir (enstrophy), and minimize the enstrophy for a given energy.  
Then, the critical point is not necessarily an equilibrium point.
In this specific problem, however, it is happens to be so, because it is the ``maximum point'' of the energy.
Notice that the minimization of the enstrophy under a constrained energy is equivalent to the maximization of the energy under a constrained enstrophy (see Appendix A).
The maximization of the energy appears to be consistent with the inverse-cascade story.
However, the simultaneous process, i.e. the forward cascade of the enstrophy violates the constancy of the enstrophy.
The dual aspects of the 2D turbulence pose a paradox in the mechanical interpretation of the selective dissipation.

The target of this study is totally different.
Whereas we formulate a variational principle using the list of ideal constants of motion,
the target functional is not selected from them.
We estimate the minimum of the enstrophy possibly given to the zonal component
(which we call the \emph{zonal enstrophy}).
Knowing how strong the zonal flow must be and how it is controlled is an important issue in the study of turbulent transport. 
While the total enstrophy is an ideal constant of motion, the zonal part alone is not.
We are not proposing that the zonal enstrophy is selectively dissipated;
we never provide the target functional with the role of dictating dissipation process
(it may be interesting to compare the minimizers of the zonal enstrophy and the total enstrophy, the target of the selective dissipation model; see Appendix B).
Our target functional is simply \emph{what we want to estimate}.
We derive an \emph{a priori} estimate of the zonal enstrophy, which must apply to every possible dynamics under a set of prescribed conditions; the ideal invariants are used as such constraints
(we do not include the adiabatic invariant, because it needs the wave number information that is not amenable to our formulation).
The actual dynamics is the second subject to be explored, which will be the task of Sec\,\ref{sec:simulation}.
The analogy of quantum mechanical energy levels may be helpful to explain our perspective.
When we want to estimate the energy of an orbital electron, the variational principle to find the critical values of energy, for a fixed total probability, leads us to the eigenvalue problem for the Hamiltonian.
The actual energy level that a particular electron will take is determined, for example,  by the deexcitation process of emitting photons.  
We will find a similar picture for the 2D-fluid turbulence;
the zonal enstrophy has discrete levels of critical values (local minimums);
by emitting wavy enstrophy, the zonal enstrophy relaxes into lower levels.

If there is no constraint on partitioning, the zonal enstrophy can be minimized to zero (even if the total enstrophy is kept at a non-zero constant).
But some constraints prevent this to occur.
We will identify the ``key constraints'' that determine the reasonable estimate of the zonal enstrophy.

The reciprocal problem, which maximizes the complementary \emph{wavy enstrophy} 
(= total enstrophy $-$ zonal enstrophy),
was first studied by Shepherd\,\citep{Shepherd} with a different motivation,
i.e. to estimate upper bounds on instabilities in nonlinear regime.
This is seemingly equivalent to the minimization of the zonal component,
however, the effective ``constraints'' may differ (see Appendix A).
The conservation of the \emph{pseudo-momentum} was invoked as the essential constraint.
Improved estimates have been proposed by taking into account more general 
set of invariants which are known as Casimirs\,\citep{IshiokaYoden}.
In the present study of the minimization of the zonal component,
however, we invoke a different constant of motion, the energy, as the principal constraint
(in addition to other ones such as angular momentum).
The physical reason is clear because the self-organization is a spontaneous process
in which the redistribution of the enstrophy between the zonal and wavy components can occur only if the energetics admits.
Moreover, the energy constraint imparts a mathematically peculiar property to the variational principle, which is the other incentive of this study.

In the next section, we will start by reviewing the basic formulation and preliminaries.
Section\,\ref{sec:estimates} describes the main result.
We will derive discrete levels of the minimum zonal enstrophy.
We will propose the notion of deexcitation to lower enstrophy levels (in analogy of energy levels of quantum states);
the relaxation into lower levels corresponds to the inverse cascade.
According to the conjecture of Rhines scale\,\citep{Rhine}, the inverse cascade continues until the linear Rossby wave term overcomes the nonlinear term.
In Sec.\,\ref{sec:simulation}, we will study the relaxation process by numerical simulation.
The conventional Rhines scale will be revisited to give an improved estimate of the relaxed zonal enstrophy level.
Section\,\ref{sec:conclusion} concludes this paper.

\section{Basic formulation and preliminaries}\label{sec:formulation}
\subsection{Vortex dynamics on a beta plane}\label{subsec:formulation}

We consider a barotropic fluid on a beta-plane
\[
M = \{\boldsymbol{\xi}=(x,y)^{\mathrm{T}} ;\, x\in[0,1), y\in (0,1) \}.
\]
Here, $x$ is the azimuthal coordinate (longitude) and $y$ is the meridional coordinate (latitude).
Identifying the points $(0,y)^{\mathrm{T}} = (1,y)^{\mathrm{T}}$, 
all functions on $M$ is periodic in $x$.
The boundary is $\partial M = \Gamma_0 \cup \Gamma_1$ with
\[
\Gamma_0 = \{\boldsymbol{\xi}=(x,0)^{\mathrm{T}} ;\, x\in[0,1) \},
\quad
\Gamma_1 = \{\boldsymbol{\xi}=(x,1)^{\mathrm{T}} ;\, x\in[0,1) \}.
\]
We will denote the standard $L^2$ inner product by $\langle f,g \rangle$:
\[
\langle f,g \rangle = \int_M f(\boldsymbol{\xi}) g(\boldsymbol{\xi}) \,\mathrm{d}^2\xi,
\]
and the $L^2$ norm by $\| f\| = \langle f, f\rangle^{1/2}$.

The state vector is the fluid vorticity $\omega \in L^2(M)$.
We define the stream function (or Gauss potential) $\psi$ by
\begin{equation}
-\Delta\psi = \omega  ,
\label{stream_function}
\end{equation}
where $\Delta=\partial_x^2 + \partial_y^2$.
The flow velocity is given by
\begin{equation}
\boldsymbol{v}=\left( \begin{array}{c} v_x \\ v_y \end{array}\right)
= \bnabla_\perp\psi
= \left( \begin{array}{c} \partial_y \psi \\  -\partial_x \psi \end{array}\right) .
\label{2Dflow}
\end{equation}
Adding a normal coordinate $z$, we embed $x$-$y$ plain in $\mathbb{R}^3$,
and consider a 3-vector $\tilde{\boldsymbol{v}} = (v_x, v_y, 0)^{\mathrm{T}}$ such  that $\partial_z \tilde{\boldsymbol{v}}=0$.
Then, we may calculate 
$\bnabla\times\tilde{\boldsymbol{v}} = (0,0,-\Delta\psi)^{\mathrm{T}} = (0,0,\omega)^{\mathrm{T}} $,
justifying that we call $\omega$ the \emph{vorticity}.

To determine $\psi$ by (\ref{stream_function}), 
we impose a homogeneous Dirichlet boundary condition
\begin{equation}
 \bigl. \psi \bigr|_{\Gamma_0}= \bigl.\psi \big|_{\Gamma_1}=0.
\label{BC1}
\end{equation}
Since $M$ is periodic in $x$, we have
\begin{equation}
\bigl. D \psi \bigr|_{x=0}= \bigl. D \psi \bigr|_{x=1},
\label{BC0}
\end{equation}
where $D$ is an arbitrary linear operator.
We note that (\ref{BC1}) implies that the flow is confined in the domain
(i.e. $\bigl.\boldsymbol{n}\bcdot\boldsymbol{v}\bigr|_{\partial M}=\bigl.v_y \bigr|_{\partial M} = 0$;
$\boldsymbol{n}$ is the unit normal vector on $\partial M$), and has zero meridian flux:
\begin{equation}
\int_0^1 v_x \,\mathrm{d} y = \int_0^1 \partial_y \psi \,\mathrm{d} y = \Bigl[ ~\psi~ \Bigr]^{y=1}_{y=0} =0.
\label{0-flux}
\end{equation}
We note that a weaker boundary condition such that
$\psi |_{\Gamma_0}=a,~ \psi |_{\Gamma_1}=b$ ($a$ and $b$ are some real constants)
maintains $v_y |_{\Gamma_0}=  v_y |_{\Gamma_1}=0$, 
but allows a finite meridian flux
(see the discussion after Proposition\,\ref{proposition:constants} of Sec.\,\ref{subsec:constants}).

We define the \emph{Laplacian} as a self-adjoint operator in $L^2(M)$ by imposing the boundary condition (\ref{BC1}) to its domain.
Its unique inverse $\mathcal{K}=(-\Delta)^{-1}$ is a compact self-adjoint operator,
by which we can solve (\ref{stream_function}) for $\psi$.

Taking into account the Coriolis force, 
the governing equation of $\omega$ is
\begin{equation}
\partial_t \omega + \{\omega + \beta y, \psi \} =0,
\label{beta-eq}
\end{equation}
where $\{ f, g \} = (\partial_x f )(\partial_y g) -  (\partial_x g)(\partial_y f) $,
and $\beta$ is a real constant number measuring the meridional variation of the Coriolis force.
When $\beta=0$, (\ref{beta-eq}) reduces into the standard vorticity equation.
A finite $\beta$ introduces anisotropy to the system, 
resulting in creation of \emph{zonal flow}. 
The Rhines scale\,\cite{Rhine} speaks of the balance of the two terms
$\{\omega , \psi \} $ and $\{ \beta y, \psi \} $,
by which we obtain the typical scale length of the zonal flow (see Sec.\,\ref{subsec:estimate_lambda})

Inverting (\ref{stream_function}) by $\mathcal{K}=(-\Delta)^{-1}$,
we may rewrite (\ref{beta-eq}) as
\begin{equation}
\partial_t \omega + \{\omega + \beta y,  \mathcal{K} \omega \} =0.
\label{beta-eq-2}
\end{equation}

We call
\begin{equation}
\omega_t := \omega + \beta y
\label{total_vorticity}
\end{equation}
the \emph{total vorticity}, which is the sum of the fluid part $\omega$ and the ambient part $\beta y$ (the latter is due to the rotation of the system).

The following identity will be useful in the later calculations:
\begin{equation}
\langle f, \{ g,h \} \rangle = \langle g, \{ h, f \} \rangle,
\label{Poisson-1}
\end{equation}
where $f, g$ and $h$ are $C^1$-class functions in $M$, and
either $f$ or $g$ satisfy (\ref{BC1}).


\subsection{Conservation laws and symmetries}
\label{subsec:constants}

\begin{proposition}[constants of motion]
\label{proposition:constants}
The following functionals are constants of motion of the evolution equation (\ref{beta-eq-2}):
\begin{enumerate}
\item
Energy:
\begin{equation}
E(\omega) :=\frac{1}{2} \langle \omega, \mathcal{K}\omega \rangle .
\label{energy}
\end{equation}
By rewriting 
\[
E= \frac{1}{2} \langle (-\Delta\psi), \psi \rangle 
=\frac{1}{2} \int_M |\bnabla\psi|^2\mathrm{d}^2 \xi
=\frac{1}{2} \int_M |\bnabla_\perp\psi|^2\mathrm{d}^2 \xi = \frac{1}{2} \int_M |\boldsymbol{v}|^2\mathrm{d}^2 \xi,
\]
we find that $E$ evaluates the kinetic energy of the flow $\boldsymbol{v}$.

\item
Longitudinal momentum:
\begin{equation}
P(\omega) := \int_M \partial_y ( \mathcal{K}\omega) \,\mathrm{d}^2 \xi,
\label{momentum}
\end{equation}
We may rewrite 
\[
P=\int_M \partial_y\psi \,\mathrm{d}^2 \xi  =\int_M v_x\,\mathrm{d}^2 \xi 
\]
to see that $P$ is the integral of the longitudinal momentum.
By (\ref{0-flux}), $P$ must be constantly zero.

\item
Circulation:
\begin{equation}
F (\omega) :=\langle 1 , \omega \rangle .
\label{circulation}
\end{equation}
Integrating by parts, we may write
\[
F = \int^1_0 \Big[ v_x \Big]^{y=1}_{y=0} \, \mathrm{d} x,
\]
which evaluates the circulation of the flow $\boldsymbol{v}$ along the boundary $\partial M$.

\item
Angular momentum:
\begin{equation}
L(\omega) :=\langle y, \omega \rangle .
\label{angular-momentum}
\end{equation}
Integrating by parts and using the boundary conditions (\ref{BC1}) and (\ref{BC0}), we may rewrite
\[
L = \int_M y (\partial_x v_y -\partial_y v_x)\,\mathrm{d}^2\xi 
= \int_M v_x\,\mathrm{d}^2\xi - \int_0^1 \Bigl[\, y v_x \,\Bigr]_{y=0}^{y=1}\,\mathrm{d} x.
\]
The first term on the right-hand side is $P$, which vanishes by (\ref{0-flux}).
Hence, 
$L$ corresponds to the angular momentum $\boldsymbol{\xi}\times\boldsymbol{v}$ averaged over the boundary.

\item
Generalized enstrophy:
\begin{equation}
G_\beta(\omega) :=\int_M f(\omega+\beta y)\,\mathrm{d}^2\xi,
\label{g-enstrophy}
\end{equation}
where  $f$ is an arbitrary $C^1$-class function,
and the argument $\omega+\beta y$ is the \emph{total vorticity} including the ambient term $\beta y$.
For $f(u)=u^2/2$, $G_\beta(\omega)$ is the conventional enstrophy of the total vorticity.

\item
Fluid enstrophy:
\begin{equation}
{Q}(\omega) := \frac{1}{2} \| \omega \|^2 .
\label{fluid-enstrophy}
\end{equation}

\end{enumerate}
\end{proposition}

\noindent
(proof)
While these conservation laws are well known, we give the proof to see how they originate.
Suppose that $\omega$ is a $C^1$-class solution of (\ref{beta-eq-2}).
Rewriting (\ref{beta-eq-2}) in terms of the total vorticity $\omega_t =\omega+\beta y$,
we have $\partial_t \omega_t + \{\omega_t, \psi \} =0$
(where $\psi = \mathcal{K}(\omega_t-\beta y)$).

(1) Using the self-adjointness of $\mathcal{K}$, we may calculate
\[
\frac{\mathrm{d}}{\mathrm{d} t} E = \langle \mathcal{K} \omega, \partial_t\omega \rangle
=  \langle\psi, \{\psi, \omega_t\} \rangle 
=  \langle \omega_t, \{ \psi,\psi \} \rangle =0.
\]

(2) To evaluate $\frac{\mathrm{d}}{\mathrm{d} t} P = \int_M (\partial_t v_x)\,\mathrm{d}^2 \xi$, 
we invoke the $x$-component of Euler's equation $\partial_t \boldsymbol{v} + (\boldsymbol{v}\bcdot\bnabla)\boldsymbol{v} = -\bnabla p +2 \boldsymbol{v}\times\boldsymbol{\Omega}$:
\[
\partial_t v_x = - v_x \partial_x v_x - v_y \partial_y v_x + \beta y v_y - \partial_x p.
\]
Integrating by parts with the boundary conditions (\ref{BC1}) and (\ref{BC0}), we observe
\begin{eqnarray*}
\frac{\mathrm{d}}{\mathrm{d} t} P &=& 
\int_M (- v_x \partial_x v_x - v_y \partial_y v_x + \beta y v_y - \partial_x p)\,\mathrm{d}^2 \xi
\\
&=&
\int_M \left[-\partial_x( v_x^2 + p) + \beta y v_y \right] \,\mathrm{d}^2 \xi
\\
&=&
-\beta \int_0^1 \Bigl[ y  \psi \Bigr]_{x=0}^{x=1}\,\mathrm{d} y =0.
\end{eqnarray*}
To derive the second line, 
we have used $\bnabla\bcdot\boldsymbol{v}=0$ to put $\partial_y v_y=-\partial_x v_x$. 

(3) Using (\ref{Poisson-1}), we obtain
\[
\frac{\mathrm{d}}{\mathrm{d} t} \langle 1, \omega \rangle =
\langle 1, \{ \psi, \omega_t\} \rangle = 
\langle \psi, \{ \omega_t,1 \} \rangle = 0.
\]

(4) Similarly we obtain
 \begin{eqnarray*}
\frac{\mathrm{d}}{\mathrm{d} t} L &=& \langle y, \{\psi, \omega + \beta y\} \rangle
= \langle y, \{\psi, \omega\} \rangle
= \langle \psi, \{ \omega, y \} \rangle
\\
&=& \int_M \psi \, \partial_x \omega \,\mathrm{d}^2\xi
= \int_M v_y \omega \,\mathrm{d}^2\xi
\\
&=& \frac{1}{2} \int_M \partial_ x (v_y^2 - v_x^2 ) \,\mathrm{d}^2\xi
=0 .
\end{eqnarray*}

(5) Using (\ref{Poisson-1}), we obtain
\[
\frac{\mathrm{d}}{\mathrm{d} t} G_\beta = \langle f'(\omega_t), \partial_t\omega_t \rangle
= \langle f'(\omega_t), \{\psi, \omega_t\} \rangle
= \langle \psi, \{ \omega_t, f'(\omega_t) \} \rangle =0.
\]

(6) The generalized enstrophy for $f(u)=u^2/2$ may be written as
\begin{eqnarray*}
G_\beta(\omega) = \frac{1}{2} \|\omega + \beta y \|^2
&=& \frac{1}{2} \|\omega\|^2 + \beta \langle y, \omega \rangle + \frac{\beta^2}{6} 
\\
&=& {Q}(\omega) + \beta L(\omega) + \frac{\beta^2}{6} 
\end{eqnarray*}
Since $G_\beta(\omega)$ and $ L(\omega)$ are constants, ${Q}(\omega)$ is also a constant.
\QED

\bigskip
Notice that $P\equiv0$ is an immediate consequence of (\ref{0-flux})
that comes form the homogeneous Dirichlet boundary condition (\ref{BC1}).
However, in the proof of the constancy of $P$ (Proposition\,\ref{proposition:constants} (3)), 
we used only $v_y|_{\Gamma_0} =v_y|_{\Gamma_1}=0$,
which may be guaranteed by a weaker boundary condition 
$\psi |_{\Gamma_0}=a,~ \psi |_{\Gamma_1}=b$ ($a$ and $b$ are some real constants).
Hence, in a more general setting of boundary condition (or the definition of $\mathcal{K}$), 
$P$ may assume a general (non-zero) constant value.
Then, a question arises:
Dose the homogeneous Dirichlet condition (\ref{BC1}) violates the generality of vortex dynamics?
The answer is no: The Galilean symmetry of the system subsumes the freedom of the foregoing $a$ and $b$.  
First, the transformation $\psi \mapsto \psi -a$ does not change $\boldsymbol{v}=\bnabla_\perp \psi$,
so we may set a generalized boundary condition to be 
$\psi |_{\Gamma_0}=0,~ \psi |_{\Gamma_1}=c$.
With $\psi_c:=c y$, we decompose $\psi = \psi_0 + \psi_c$ so that $\psi_0$ satisfies the homogenized boundary condition (\ref{BC1}).
We have $\bnabla_\perp \psi_c = c \bnabla x$, a constant velocity in the longitudinal direction, and
$\omega = -\Delta \psi = -\Delta \psi_0$.
Inserting this into (\ref{beta-eq}), we obtain
\[
\partial_t \omega +\{ \omega +\beta y, \psi_0\} + c \{\omega, y\} =0,
\]
The distraction $c \{\omega, y\} = c \partial_x \omega$
can be cleared by Galilean boost $x \mapsto x - ct$.
In the inertial frame, we may put $\psi_0 = \mathcal{K} \omega$ to reproduce (\ref{beta-eq-2}).

Evidently, we have
\begin{lemma}[translational symmetry]
The constants of motion $E$, $P$, $F$, $L$, $G_\beta$, and $W$ are invariant against the transformation
\begin{equation}
T(\tau) :\,\omega(x,y) \mapsto \omega(x+\tau ,y),
\quad (\tau\in\mathbb{R}).
\label{transformation}
\end{equation}
\label{lemma:symmetry}
\end{lemma}

\section{Zonal and wavy components}\label{sec:projectors}
The phase space of the vorticity $\omega$ is 
\begin{equation}
{V} = L^2(M).
\label{phase_space}
\end{equation}
We say that $\omega$ is \emph{zonal} when $\partial_x \omega \equiv 0$ in $M$.
The totality of zonal flows deifies a closed subspace ${V}_z \subset{V}$.
The \emph{zonal average} 
\begin{equation}
\mathcal{P}_z \omega := \int_0^1 {\omega}(x,y)\,\mathrm{d} x
\label{projector}
\end{equation}
may be regarded as a projection from ${V}$ onto ${V}_z$. 
The following basic properties of the projector $\mathcal{P}_z$ may be known to the reader, but we summarize them as Lemmas for the convenience of the analysis in Sec.\,\ref{sec:estimates}.
By the orthogonal decomposition $V = V_z \oplus V_w$, we define the orthogonal complement
${V}_w$, i.e., ${\omega}_w \in{V}_w$, iff
$ \langle {\omega}_w, \omega_z\rangle = 0$ for all $\omega_z\in {V}_z$.
We call ${\omega}_w \in {V}_w$ a \emph{wavy} component, which has zero zonal average:
$\mathcal{P}_z \omega_w = 0$.
We will denote
\[
\mathcal{P}_w = I- \mathcal{P}_z ,
\]
which is the projector onto $V_w$.  
Now we may write 
\[
V = V_z \oplus V_w = (\mathcal{P}_z V) \oplus (\mathcal{P}_w V).
\]
Being projectors, $\mathcal{P}_z$ and $\mathcal{P}_w$ satisfy
$\mathcal{P}_z\mathcal{P}_z=\mathcal{P}_z$,
$\mathcal{P}_w\mathcal{P}_w=\mathcal{P}_w$, and
$\mathcal{P}_z\mathcal{P}_w=\mathcal{P}_w\mathcal{P}_z=0$.
We also have the following useful identity:

\begin{lemma}[commutativity]
\label{lemma:commutation}
Let $M$ be a beta-plane (which is periodic in $x$).
For $\psi  \in \mathcal{K}(V)$, we have
\begin{equation}
\mathcal{P}_z \Delta \psi = \Delta \mathcal{P}_z \psi .
\label{commutation1}
\end{equation}
For $\omega \in V$, we have
\begin{eqnarray}
\mathcal{P}_z \mathcal{K} \omega &=& \mathcal{K} \mathcal{P}_z \omega ,
\label{commutation2}
\\
\mathcal{P}_w \mathcal{K} \omega &=& \mathcal{K} \mathcal{P}_w \omega .
\label{commutation3}
\end{eqnarray}
\end{lemma}
 
 \noindent
(proof)
By the periodicity in $x$, we may calculate as
\[
\mathcal{P}_z \Delta \psi = \int_0^1 (\partial_x^2 \psi +\partial_y^2 \psi )\,\mathrm{d} x
=  \Bigl[ \partial_x \psi \Big]_{x=0}^{x=1} + \partial_y^2 \int_0^1 \psi\,\mathrm{d} x
= \Delta \mathcal{P}_z \psi.
\]
Putting $\psi=\mathcal{K}\omega$, (\ref{commutation1}) reads
$-\mathcal{P}_z \omega = \Delta \mathcal{P}_z \mathcal{K}\omega$.
Operating $\mathcal{K}$ on both sides yields (\ref{commutation2}).
Using this, we obtain
$\mathcal{P}_w\mathcal{K} \omega = (1-\mathcal{P}_z)\mathcal{K}\omega 
= \mathcal{K}(1 -\mathcal{P}_z)\omega
= \mathcal{K}\mathcal{P}_w\omega$.
\QED

The following properties are useful:

\begin{lemma}[partition laws]
\label{lemma:partition}
Let us decompose $\omega = \omega_z + \omega_w
~(\omega_z = \mathcal{P}_z \omega \in V_z,~ \omega_w = \mathcal{P}_w\omega \in V_w)$.
\begin{enumerate}
\item
The circulation is occupied by the zonal component $\omega_z$, i.e.,
\begin{equation}
F(\omega) = F(\omega_z).
\label{partition-F}
\end{equation}
\item
The angular momentum is occupied by the zonal component $\omega_z$, i.e.,
\begin{equation}
L(\omega) = L(\omega_z).
\label{partition-L}
\end{equation}
\item
The fluid enstrophy is simply separated as
\begin{equation}
{Q}(\omega) = {Q}(\omega_z) + {Q}(\omega_w).
\label{partition-W}
\end{equation}

\item
The energy is simply separated as
\begin{equation}
E(\omega) = E(\omega_z) + E(\omega_w) .
\label{partition-E}
\end{equation}
\end{enumerate}
\end{lemma}

\noindent
(proof)
The first three relations are clear.  
The energy partition (\ref{partition-E}) is due to
\[
\langle \omega_z , \mathcal{K} \omega_w \rangle =
\langle \omega_w , \mathcal{K} \omega_z \rangle =0,
\]
which follows from (\ref{commutation2}).
\QED

\bigskip

Evidently, $\partial_x (\mathcal{K}{\omega}_z)=0$ for $\omega_z \in V_z$.
Hence, $\{ \omega_z + \beta y, \mathcal{K}\omega_z\}=0$,
implying that every member $\omega_z \in V_z$ is a stationary solution of (\ref{beta-eq-2}).

\section{Estimate of zonal enstrophy}\label{sec:estimates}
\subsection{Zonal enstrophy vs. wavy enstrophy}\label{subsec:W}

The aim of this work is to find the minimum of the \emph{zonal enstrophy} defined by
\begin{equation}
Z(\omega) := \frac{1}{2} \| \mathcal{P}_z \omega  \|^2 .
\label{zonal-enstrophy}
\end{equation}
The complementary \emph{wavy enstrophy} is ${W}(\omega) = \frac{1}{2} \| \mathcal{P}_w \omega  \|^2 $.
By (\ref{partition-W}), the total enstrophy is 
\[
{Q}(\omega)  = {Q}(\mathcal{P}_z\omega) + {Q}(\mathcal{P}_w\omega) = Z(\omega) + {W}(\omega).
\]
When the total enstrophy ${Q}(\omega)$ is conserved
(see Proposition\,\ref{proposition:constants}\,(5)),
the minimum of $Z(\omega)$ gives the maximum of ${W}(\omega)$.

The simplest version of the minimization problem is to find the minimum $Z(\omega)$ under the constraint of ${Q}(\omega)=C_Q ~(\neq0)$.
Introducing a Lagrange multiplier $\nu$, we minimize
\begin{equation}
Z(\omega)-\nu {Q}(\omega) .
\label{Z_W}
\end{equation}
Using the self-adjointness of $\mathcal{P}_z$, we obtain the 
the Euler-Lagrange equation 
\begin{equation}
\mathcal{P}_z \omega -\nu \omega =0.
\label{EL-eq-W}
\end{equation}
Operating $\mathcal{P}_z$ on (\ref{EL-eq-W}) yields 
\[
(1-\nu) \mathcal{P}_z\omega =0.
\]
On the other hand, operating $\mathcal{P}_w$ yields
\[
\nu \mathcal{P}_w\omega =0.
\]
There are two possibilities of solving these simultaneous equations.
\begin{enumerate}
\item
$\nu=0$:  
Then, $\mathcal{P}_z\omega =\omega_z=0$ and $\mathcal{P}_w\omega =\omega_w$ is an arbitrary function
satisfying ${Q}(\omega_w)=C_Q$; hence, $\mathrm{min}\,Z(\omega)=0$.
\item
$\nu=1$:
Then, $\mathcal{P}_w\omega =\omega_w=0$ and $\mathcal{P}_z\omega =\omega_z$ is an arbitrary function
satisfying ${Q}(\omega_z)=Z(\omega)=C_Q$;
hence, this solution gives the ``maximum'' of $Z(\omega)$.
\end{enumerate}

This simple exercise reveals an unusual aspect of the present variational principle,
which is caused by the \emph{non-coerciveness} of the functional $Z(\omega)$ to be minimized.
Notice that the minimizer is not unique, because $\mathcal{P}_z$ has nontrivial kernel, i.e.
$\mathrm{Ker}(\mathcal{P}_z) = V_w$;
every $\omega_w\in V_w$ satisfies (\ref{EL-eq-W}).

To obtain a nontrivial estimate of the minimum $Z(\omega)$, 
we have to take into account ``constraints'' posed on the dynamics of redistributing enstrophy.
Guided by Proposition\,\ref{proposition:constants}, we start with some simple ones.

\subsection{Constraints by circulation and angular momentum}\label{subsec:W-F-L}

Let us consider the circulation and angular momentum as constraints.

\begin{theorem}
\label{prop:linear}
The minimizer of the zonal enstrophy $Z(\omega)$ 
under the constraints on the circulation $F(\omega)=C_F$,
the angular momentum $L(\omega)=C_L $,
as well as the total enstrophy ${Q}(\omega)=C_Q$ is a vorticity $\omega$ such that
\begin{equation}
\mathcal{P}_z \omega = a + b y ,
\quad
(a =4C_F - 6 C_L,~ b=12 C_L - 6 C_F),
\label{EL-eq-LW}
\end{equation}
which gives
\begin{equation}
Z_0 := \mathrm{min}\,Z(\omega) = 2C_F^2 - 6 C_F C_L + 6 C_L^2.
\label{min_Z_under_F_L_const}
\end{equation}
\end{theorem}

\noindent
(proof) 
Let us minimize 
\begin{equation}
Z(\omega) - \nu {Q}(\omega) - \mu_0 F(\omega)-\mu_1 L(\omega).
\label{Z_W-F-L}
\end{equation}
The Euler-Lagrange equation is
\begin{equation}
\mathcal{P}_z \omega - \nu \omega =\mu_0 +  \mu_1 y.
\label{EL-eq-LW-1}
\end{equation}
Operating $\mathcal{P}_z$ on both sides of (\ref{EL-eq-LW-1}) yields 
\begin{equation}
(1-\nu) \mathcal{P}_z\omega =\mu_0 +  \mu_1 y.
\label{EL-eq-LW-1'}
\end{equation}
On the other hand, operating $\mathcal{P}_w$ yields 
\begin{equation}
\nu \mathcal{P}_w\omega=0.
\label{EL-eq-LW-2}
\end{equation}
First, assume that $1-\nu\neq 0$.
Inserting $\mathcal{P}_z\omega$ of (\ref{EL-eq-LW-1'}) into the definition of $F(\omega)=F(\mathcal{P}_z\omega)$
and $L(\omega)=L(\mathcal{P}_z\omega)$
(see Lemma\,\ref{lemma:partition} (1) and (2)),
we determine $\mu_0$ and $\mu_1$ of to match the constraint
 $\langle 1 , \omega\rangle ={C_F}$ and 
 $\langle y , \omega\rangle ={C_L}$;
we obtain
$a:= \mu_0/(1-\nu)=4C_F - 6 C_L$, and $b:=\mu_1/(1-\nu) = 12 C_L - 6 C_F$.
Inserting this $\omega_z = a + b y$ into $Z(\omega)$, we obtain the minimum
(\ref{min_Z_under_F_L_const}).
On the other hand, (\ref{EL-eq-LW-2}) is satisfied by $\nu =0$ (consistent with the forgoing assumption  $1-\nu\neq 0$) and an arbitrary $\omega_w = \mathcal{P}_w\omega$ such that
\begin{equation}
\frac{1}{2} \| \omega_w\|^2 = C_Q -(2C_F^2 - 6 C_F C_L + 6 C_L^2) .
\label{omega_w_of_L}
\end{equation}
The right-hand side is non-negative, if the constraints $F(\omega)=C_F$, 
$L(\omega)={C_L}$ and ${Q}(\omega)=C_Q$ are consistent.
It is only when the constants $C_F$, $C_L$ and $C_Q$ are given so that the right-hand side of (\ref{omega_w_of_L}) is zero, 
that the other assumption $1-\nu=0$ applies;
then, the unique solution $\mathcal{P}_w\omega=0$ (hence, $\omega=\mathcal{P}_z\omega$) is obtained.
\QED

Notice that the minimizer is still non-unique (excepting the special case mentioned in the proof);
every $a + b y + \omega_w$ ($\forall \omega_w\in V_w$ such that (\ref{omega_w_of_L}) holds) 
satisfies (\ref{EL-eq-LW}).
However, the minimum value (\ref{min_Z_under_F_L_const}) 
is uniquely determined.

\subsection{Constraint by energy}
\label{subsec:W-E}
The situation changes dramatically, when we include the energy constraint $E(\omega)=C_E$;
laminated vorticity distribution, epitomizing the structure of zonal flow,
is created by the energy constraint.
The number of lamination (jet number) is identified by the ``eigenvalue'' of the Euler-Lagrange equation,
which specifies the ``level'' of the zonal enstrophy 
(in analogy of the quantum number of discrete energy in quantum mechanics).
To highlight the role of the energy constraint, we first omit the constraints on the circulation and angular momentum.

\begin{figure} [H]
\begin{center}
\includegraphics[scale=0.3]{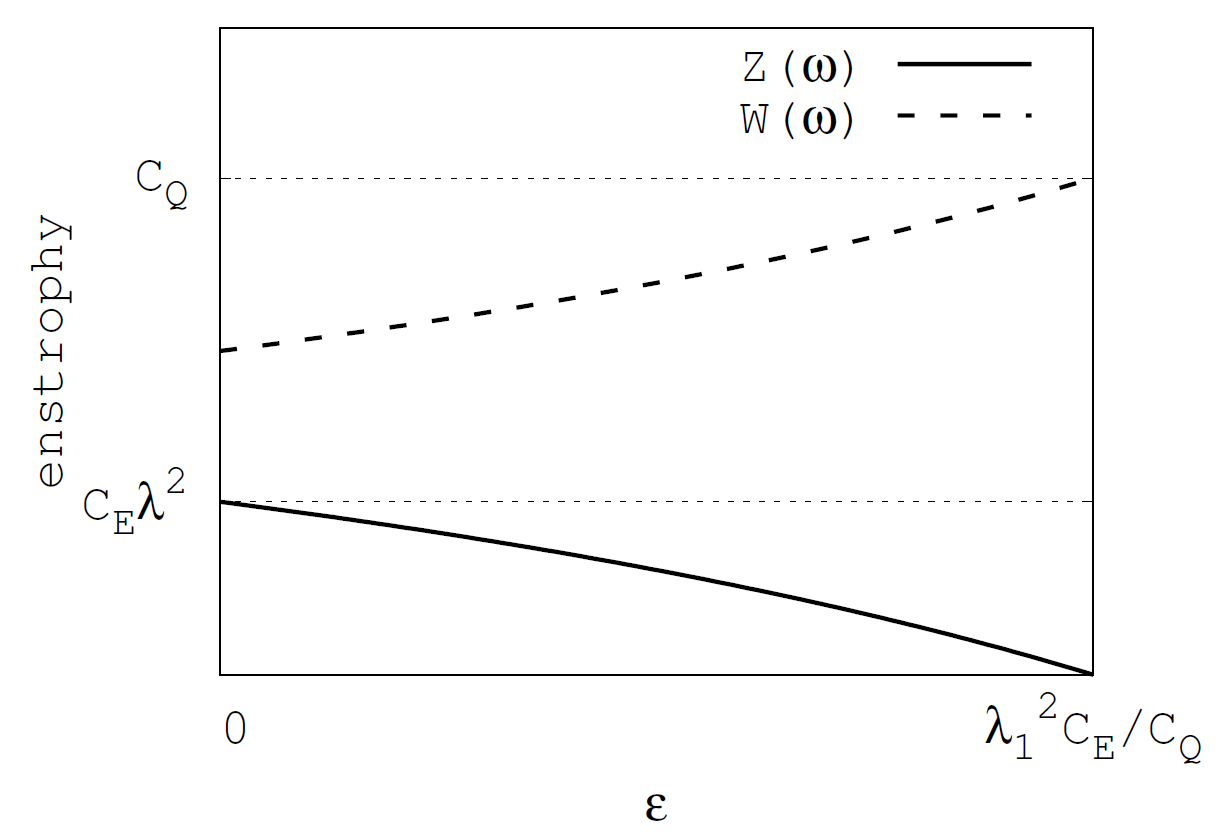}
\end{center}
\caption{
\label{fig:graph1}
The graph of $Z(\omega)$ and $W(\omega)$ given in (\ref{Z-estimate1}) and (\ref{W_omega-estimate1}).
}
\end{figure}

Taking into account the energy and total enstrophy constraint, we seek the critical points of
\[
Z(\omega) - \nu Q (\omega) - \mu_2 E(\omega).
\]
The Euler-Lagrange equation is
\begin{equation}
\mathcal{P}_z \omega - \nu \omega - \mu_2 \mathcal{K}\omega = 0.
\label{ELeq-7}
\end{equation}
Operating $\mathcal{P}_z$ yields (denoting $\omega_z = \mathcal{P}_z\omega$)
\begin{equation}
\omega_z - \nu \omega_z - \mu_2 \mathcal{K}\omega_z = 0.
\label{ELeq-P1}
\end{equation}
On the other hand, $\omega_w = \mathcal{P}_w\omega$ must satisfy
\begin{equation}
\nu \omega_w + \mu_2 \mathcal{K}\omega_w = 0.
\label{ELeq-Q1}
\end{equation}
Putting $\omega_z = -\partial_y^2 \psi_z(y)$ in (\ref{ELeq-P1}), we obtain
\begin{equation}
\partial_y^2 \psi_z + \lambda^2 \psi_z = 0 ,
\quad \lambda^2 =\frac{\mu_2}{1-\nu} .
\label{ELeq-8}
\end{equation}
The solution satisfying the boundary conditions $\psi_z(0)=\psi_z(1)=0$ is  
\begin{equation}
\psi_z = A \sin \lambda y  
\label{zonal-potential1}
\end{equation}
with eigenvalues
\[
\lambda = n_1 \pi \ (n_1 \in \mathbb{Z}).
\]
The corresponding zonal vorticity is
\begin{equation}
\omega_z = A \lambda^2 \sin \lambda y .
\label{ELeq-9}
\end{equation}
On the other hand, putting $\omega_w = -\Delta \psi_w$, (\ref{ELeq-Q1}) reads 
\begin{equation}
\Delta \psi_w + k^2 \psi_w = 0 ,
\quad k^2 = -\frac{\mu_2}{\nu}.
\label{ELeq-10}
\end{equation}
The solution satisfying the boundary conditions $\psi_w(x, 0)=\psi_w(x, 1)=0$,
as well as the periodicity in $x$, is given by (as the equivalent class of the translational symmetry in $x$;
see Lemma\,\ref{lemma:symmetry})
\begin{equation}
\psi_w = B \sin k_x x \ \sin k_y y  ,
\quad k^2=k_x^2+k_y^2
\label{wave-potential1}
\end{equation}
with eigenvalues
\[
k_x = 2 n_2 \pi , \ k_y = n_3 \pi \ (n_2,\ n_3 \in \mathbb{Z}) .
\]
The corresponding wavy vorticity is
\begin{equation}
\omega_w = B k^2 \sin k_x x \, \sin k_y y .
\label{ELeq-11}
\end{equation}
Summing the zonal and wavy components, we obtain
\begin{equation}
\psi = A \sin \lambda y + B \sin k_x x \ \sin k_y y ,
\label{potential1}
\end{equation}
\begin{equation}
\omega = A \lambda^2 \sin \lambda y +  B k^2 \sin k_x x \ \sin k_y y .
\label{vorticity1}
\end{equation}
The two amplitudes $A$ and $B$ are determined by the constraints $E(\omega)=C_E$ and $Q(\omega)=C_Q$;
inserting (\ref{potential1}) and (\ref{vorticity1}) into the definitions of $E(\omega)$ and $Q(\omega)$, we obtain
\begin{equation}
C_E = \frac{A^2 \lambda^2}{4} +  \frac{B^2 k^2}{8} ,
\label{C_E-estimate1}
\end{equation}
\begin{equation}
C_Q = \frac{A^2 \lambda^4} {4}+ \frac{B^2 k^4} {8}.
\label{C_W-estimate1}
\end{equation}
Solving (\ref{C_E-estimate1}) and (\ref{C_W-estimate1}) for $A$ and $B$, and
inserting the solution into the zonal enstrophy $Z(\omega)$ and wavy enstrophy $W(\omega)$,
we obtain the critical values
\begin{eqnarray}
Z_{\lambda,\epsilon} &=& \frac{\lambda^2 C_E -\epsilon C_Q } {1 -\epsilon},
\label{Z-estimate1}
\\
W_{\lambda,\epsilon} &=& \frac{C_Q - \lambda^2 C_E} {1 - \epsilon},
\label{W_omega-estimate1}
\end{eqnarray}
where $\epsilon = \lambda^2/k^2$, 
scaling the ratio of the wave length of the zonal components to that of the wavy components.
For $Z_{\lambda,\epsilon}\geq 0$ and $W_{\lambda,\epsilon}\geq 0$, 
there are two possibilities: $\epsilon \leq (\lambda^2 C_E)/C_Q \leq 1$ or 
$\epsilon \geq (\lambda^2 C_E)/C_Q \geq 1$.
Here, the former regime of $\epsilon$ is relevant, because we assume that the wavy components have smaller scales in comparison with the zonal component (i.e. $\epsilon<1$).
Then, $Z_{\lambda,\epsilon}$ of (\ref{Z-estimate1}) increases monotonically as $\epsilon$ decreases
(or $k^2$ increases; see Fig.\,\ref{fig:graph1}), and we have
\begin{equation}
\lim_{\epsilon \to 0} Z_{\lambda,\epsilon} = \lambda^2 C_E .
\label{Z-max}
\end{equation}
Notice that this limit gives the upper bound for $Z(\omega) $ of the corresponding eigenvalue $\lambda$, which is achieved when the wavy component has the smallest scale $\epsilon\rightarrow0$.
For actual wavy components, $Z(\omega) $ takes a smaller value than $\lambda^2 C_E$, i.e.
\begin{equation}
Z(\omega) \leq \lambda^2 C_E .
\label{Z-max-2}
\end{equation}

\subsection{Constraints by energy, circulation, angular momentum and total enstrophy}
\label{subsec:W-E-L-F}

Now we study the minimum of the zonal enstrophy $Z(\omega)$ under the all  
constraints of energy, circulation, angular momentum, and total enstrophy. 
In contrast to the observation of Sec.\,\ref{subsec:W-E} 
(where the minimum of $Z(\omega)$ is not determined by the energy $C_E$),
we will find that the minimum of $Z(\omega)$ is determined by the circulation $C_F$ and angular momentum $C_L$.
In comparison with the result of Sec.\,\ref{subsec:W-F-L}, however, we have a discrete set of enstrophy levels (each of them corresponds to different lamination number of zonal flow).
Whereas they are due to the energy constraint,
$Z(\omega)$ itself does not depend on the values of the energy $C_E$.

Introducing Lagrange multipliers, we seek the minimizer of
\[
Z(\omega) - \nu Q(\omega) - \mu_0 F(\omega) -\mu_1 L(\omega) - \mu_2 E(\omega).
\]
The Euler-Lagrange equation is
\begin{equation}
\mathcal{P}_z \omega - \nu \omega -\mu_0 -\mu_1 y - \mu_2 \mathcal{K}\omega = 0.
\label{ELeq-12}
\end{equation}
The solution satisfying the boundary conditions $\psi(x,0)=\psi(x,1)=0$,
as well as the periodicity in $x$, is $\psi=\psi_z + \psi_w$ with
\begin{eqnarray}
\psi_z &=& A_1 \cos \lambda y + A_2 \sin \lambda y - \frac{\mu_0 + \mu_1 y}{\mu_2} ,
\label{zonal-potential2}
\\
\psi_w &=& B \sin k_x x \ \sin k_y y  ,
\label{wave-potential2}
\end{eqnarray}
where
\begin{equation}
\lambda=\sqrt{\frac{\mu_2}{1-\nu}},
\quad
k^2 =k_x^2 +k_y^2 = -\frac{\mu_2}{\nu} ,
\label{lambda_zonal}
\end{equation}
and 
\begin{equation}
k_x = 2 n_2 \pi , \ k_y = n_3 \pi \ (n_2,\ n_3 \in \mathbb{Z}).
\label{k's_zonal}
\end{equation}
The corresponding vorticities are
\begin{eqnarray}
\omega_z &=& A_1 \lambda^2 \cos \lambda y + A_2 \lambda^2 \sin \lambda y.
\label{ELeq-13}
\\
\omega_w &=& B k^2 \sin k_x x \ \sin k_y y .
\label{ELeq-14}
\end{eqnarray}
The zonal enstrophy $Z(\omega)$ of the minimizer is
\begin{eqnarray}
Z(\omega) &=& \frac{A_1^2 \lambda^3}{8} (2 \lambda + \sin 2 \lambda) + \frac{A_2^2 \lambda^3}{8} (2 \lambda - \sin 2 \lambda) 
\nonumber
\\
&&~+  \frac{A_1 A_2 \lambda^3}{4} (1 - \cos 2 \lambda).
\label{Z-estimate2}
\end{eqnarray}

We have yet to determine the eigenvalue $\lambda$ and the coefficients $A_1$, $A_2$ and $B$.
Inserting $\psi=\psi_z+\psi_w$ and $\omega=\omega_z+\omega_w$ into the constraints $F(\omega)=C_F$,  $L(\omega)=C_L$,  $E(\omega)=C_E$, and $Q(\omega)=C_Q$,
we obtain
\begin{eqnarray}
C_F &=&A_1 \lambda \sin \lambda + A_2 \lambda (1- \cos \lambda) ,
\label{C_F-estimate2}
\\
C_L &=& A_1 (\lambda \sin \lambda+ \cos \lambda- 1) + A_2 (\sin \lambda- \lambda \cos \lambda) ,
\label{C_L-estimate2}
\\
C_E &=& \frac{A_1^2 \lambda}{8} (2 \lambda + \sin 2 \lambda) + \frac{A_2^2 \lambda}{8} (2 \lambda - \sin 2 \lambda) 
\nonumber 
\\
& & + \frac{A_1 A_2 \lambda}{4} (1 - \cos 2 \lambda)
- \frac{A_1 C_F}{2} 
\nonumber
\\
& & - \frac{[A_1 (\cos \lambda - 1) + A_2 \sin \lambda] C_L}{2} + \frac{B^2 k^2}{8},
\label{C_E-estimate2}
\\
C_Q &=& \frac{A_1^2 \lambda^3}{8} (2 \lambda + \sin 2 \lambda) + \frac{A_2^2 \lambda^3}{8} (2 \lambda - \sin 2 \lambda) 
\nonumber
\\
&&  +  \frac{A_1 A_2 \lambda^3}{4} (1 - \cos 2 \lambda) + \frac{B^2 k^4}{8}.
\label{C_W-estimate2}
\end{eqnarray}
We may write (\ref{C_F-estimate2}) and (\ref{C_L-estimate2}) as
\begin{equation}
\setlength{\arraycolsep}{0pt}
\renewcommand{\arraystretch}{1.3}
\left(
\begin{array}{c}
C_F \\ 
\displaystyle
C_L
\end{array}  \right)
=
\mathsfbi{D}(\lambda)
\left(
\begin{array}{c}
A_1 \\ 
A_2
\end{array}  \right) .
\label{C_F-C_L}
\end{equation}
with
\begin{equation}
\mathsfbi{D}(\lambda)
:=
\setlength{\arraycolsep}{10pt}
\renewcommand{\arraystretch}{1.3}
\left(
\begin{array}{cc}
\lambda \sin \lambda &\lambda (1- \cos \lambda) \\ 
\displaystyle
\lambda \sin \lambda+ \cos \lambda- 1 &\sin \lambda- \lambda \cos \lambda
\end{array}  \right) .
\end{equation}
For given $C_F$ and $C_L$, we solve (\ref{C_F-C_L}) to determine the amplitudes of zonal vorticity:
\begin{eqnarray}
A_1 &=& \frac{C_F (\sin \lambda - \lambda \cos \lambda) + C_L (-\lambda + \lambda \cos \lambda)}{\mathrm{det}{D}(\lambda)},
\label{A_1}
\\
A_2 &=& \frac{C_F (-\lambda \sin \lambda - \cos \lambda + 1) + C_L \lambda \sin \lambda}{\mathrm{det}{D}(\lambda)},
\label{A_2}
\end{eqnarray}
where $\mathrm{det}{D}(\lambda) = \lambda (2 - \lambda \sin \lambda - 2 \cos \lambda)$. 
Inserting (\ref{A_1}) and (\ref{A_2}) into (\ref{Z-estimate2}), we obtain the zonal enstrophy evaluated as a function of $\lambda$,
which we denote by $Z_\lambda$. 
The critical points (local minimums) of $Z(\omega)$, given by
\begin{equation}
\frac{\mathrm{d}}{\mathrm{d}\lambda}Z_\lambda = 0,
\label{eigenvalue_alpha}
\end{equation}
determine the eigenvalues $\lambda$ characterizing the \emph{enstrophy levels}. 

\begin{figure} [H]
\begin{center}
\includegraphics[scale=0.3]{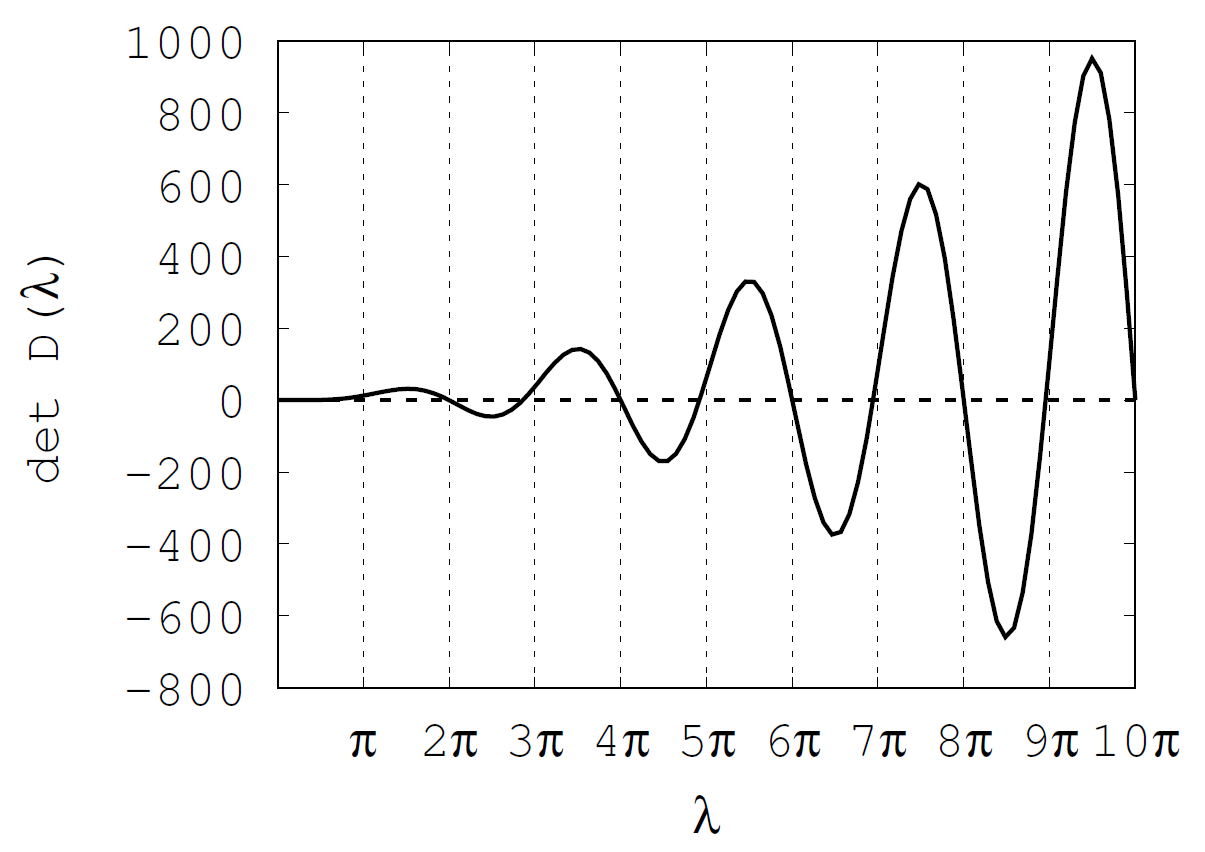}
\end{center}
\caption{
\label{fig:graph2}
The graph of $\mathrm{det}{D}$.
}
\end{figure}

Instead of displaying the lengthy expression of $Z_\lambda$, we will show its graphs for typical choices of the parameters $C_F$ and $C_L$.
Notice that $Z_\lambda$ depends only on $C_F$ (circulation) and $C_L$ (angular momentum); 
it does not contain $C_E$ (energy) and $C_Q$ (enstrophy) as parameters.
First, we pay attention to the singularities given by $\mathrm{det}{D}(\lambda)=0$,
where $A_1 \to \infty$ and $A_0 \to \infty$, hence $Z_\lambda \to \infty$
(there is an exception, as discussed later). 
We show the graph of $\mathrm{det}{D}(\lambda)$ in Fig.\,\ref{fig:graph2}. 


There are two types of solutions: 
\[
\lambda= \left\{ \begin{array}{l}
\Lambda_{2n}=2 n \pi ,
\\
\Lambda_{2n+1}= (2n+1)\pi - \delta_n ,
\end{array} \right.
\quad (n=0, 1, \cdots),
\]
where each $\delta_n$ is a small positive number such that $\delta_n \rightarrow 0$ as $n\rightarrow\infty$.
The minimums of $Z_\lambda$ appear in every interval $(\Lambda_{2n}, \Lambda_{2n+1})$.
However, if $C_F = 2 C_L$, $Z_\lambda$ remains finite at $\lambda = \Lambda_{2n+1}$.
In this special case, the minimums of $Z_\lambda$ appear in intervals $(\Lambda_{2n}, \Lambda_{2n+2})$.

\begin{figure} [H]
\begin{center}
\includegraphics[scale=0.45]{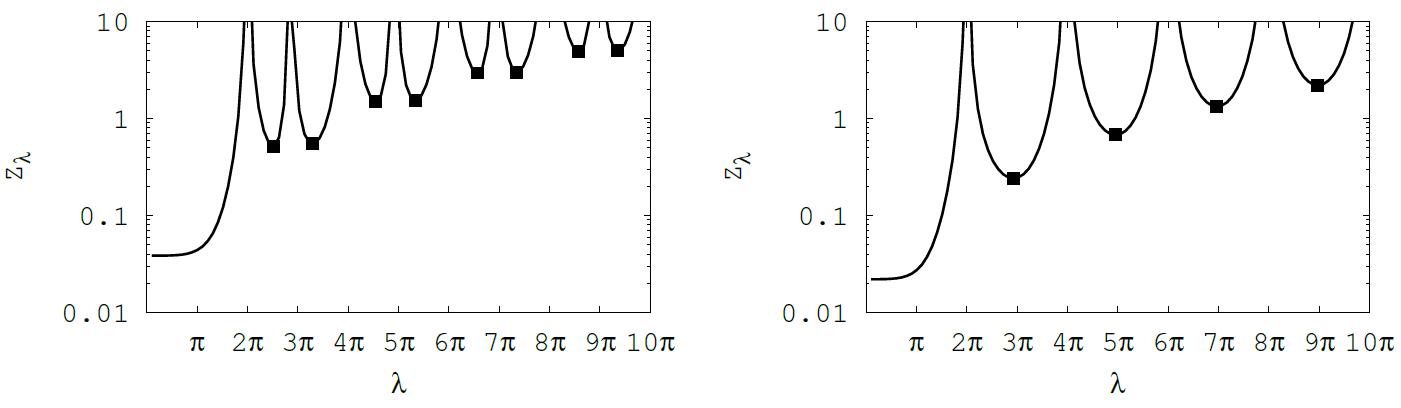}
\end{center}
\caption{
\label{fig:graph3}
The graphs of the critical zonal enstrophy $Z_\lambda$ as functions of $\lambda$.
The minimums of $Z_\lambda$ determine the eigenvalues of $\lambda$. 
The points on the curve indicates the eigenvalues.
We assume (left) $C_F=0.21$ and $C_L=0.0525$, and (right) $C_F=0.21$ and $C_L=0.105$.
}
\end{figure}

In Fig.\,\ref{fig:graph3}, we show examples of $Z_\lambda$ calculated for
 (left) $C_F=0.28$ and $C_L=0.07$,
 (right) $C_F=0.28$ and $C_L=0.14$ ($C_F = 2 C_L$).


At $\lambda =0$, $Z_\lambda$ reproduces the result of Theorem\,\ref{prop:linear},
i.e.
\begin{equation}
\lim_{\lambda \to 0} Z_\lambda = Z_0=2C_F^2 - 6 C_F C_L + 6 C_L^2,
\end{equation}
which is the absolute minimum of the zonal enstrophy under the constraints
on the circulation $F(\omega)=C_F$, the angular momentum $L(\omega)=C_L $, and the total enstrophy ${Q}(\omega)=C_Q$.

The role of the energy constraint $E(\omega)=C_E$ is to create eigenvalues of $\lambda$ at which $Z_\lambda$ takes \emph{local minimum} values.
However, the value of $C_E$ does not influence the value of $Z_\lambda$ directly.
As we have seen in (\ref{Z-max-2}),
it poses a constraint on the maximum:
\begin{equation}
Z(\omega) \leq \lambda^2 C_E,
\label{constraint_by_C-E}
\end{equation}
in addition to the other implicit constraint $Z(\omega)\leq C_Q$.
Instead of the zonal component $\omega_z$ of (\ref{ELeq-13}), $C_E$ and $C_Q$ work for determining the complementary wavy component $\omega_w$ of (\ref{ELeq-14}).
By (\ref{C_E-estimate2}) and (\ref{C_W-estimate2}), we obtain 
\begin{eqnarray}
k^2 &=& \frac{C_Q-Z_\lambda}{C_E-E_{z,\lambda}} ,
\label{k_lambda}
\\
B^2 &=& \frac{8(C_E-E_{z,\lambda})^2}{C_Q-Z_\lambda},
\label{B_lambda}
\end{eqnarray}
where $E_{z,\lambda}$ is the energy of the zonal component $\omega_z$ evaluated at the eigenvalue $\lambda$.
Notice that $k^2 B^2$ ($\sim$ energy of the wavy component) is determined only by $C_E$ and  $E_{z,\lambda}$.
So, the role of the total enstrophy constraint is to determine the wave number $k$ of the wavy component.


For the special case of $C_F=0$ and $C_L=0$, a laminated zonal flow ($A_1\neq0$ and/or $A_2\neq0$) can occur only if 
\[
\mathrm{det}{D}(\lambda) = \lambda (2 - \lambda \sin \lambda - 2 \cos \lambda) = 0.
\]
Then, the eigenvalues are 
$\lambda=\Lambda_{2n}$ and  $\Lambda_{2n+1}$ ($n=0,1,2,\cdots$), the previous singular points;
see Fig.\,\ref{fig:graph2}.
For $\lambda=\Lambda_{2n}$
($\lambda=0$ gives the trivial solution $\omega_z=0$),
\[
\mathsfbi{D}(\lambda) =
\renewcommand{\arraystretch}{1.3}
\left(
\begin{array}{cc}
0 & 0 \\ 
0 & - \lambda
\end{array}  \right),
\]
hence, $A_2=0$.
On the other hand, for $\lambda=\Lambda_{2n+1}$, 
\[
\mathsfbi{D}(\lambda) =
\frac{1}{4} \lambda \sin \lambda
\renewcommand{\arraystretch}{1.3}
\left(
\begin{array}{cc}
4 & 2 \lambda \\ 
2 & \lambda
\end{array}  \right) ,
\]
and then $A_2 = -2 A_1/\Lambda_{2n}$.
In both cases, $A_1$ is arbitrary, so 
we cannot determine the amplitude of the zonal vorticity $\omega_z$.
Therefore, the trivial conditions $C_F=0$ and $C_L=0$ reproduce the situation of ``no-constraint'' discussed in Sec.\,\ref{subsec:W-E}.
We only have the estimate of the maximum (\ref{Z-max-2}).

The forgoing results are summarized as:
\begin{theorem}
\label{theorem:laminated}
For a given set of constants
$F(\omega)=C_F$, $L(\omega)=C_L $,
$E(\omega)=C_E$, and ${Q}(\omega)=C_Q$,
the zonal enstrophy $Z(\omega)$ has a discrete set of critical (local minimum) values 
quantized by the eigenvalue $\lambda$ measuring the lamination period of the zonal vorticity.
\begin{enumerate}
\item
When $C_F\neq0$ or $C_L\neq0$, the eigenvalue $\lambda$ is given by (\ref{eigenvalue_alpha}) as a function of $C_F$ and $C_L$.
The corresponding eigenfunction $\omega_z$, and 
the critical value of $Z(\omega)$ are determined by $C_F$ and $C_L$; see (\ref{ELeq-13}), (\ref{Z-estimate2}), (\ref{A_1}) and (\ref{A_2}). 
The other constants $C_E$ and $C_Q$ determine upper bounds  $C_E \lambda^2 \geq Z(\omega)$ and $C_Q \geq Z(\omega)$.
\item
For the special values $C_F=C_L=0$,  additional eigenvalues $\lambda=2n\pi$ and $\lambda=\Lambda_n$ ($n=1,2,\cdots$) occur.
However, the eigenfunctions $\omega_z$ and the critical values of $Z(\omega)$ are no longer determined by such $C_F$ and $C_L$;
we only have estimates of upper bounds $C_E \lambda^2 \geq Z(\omega)$ and $C_Q \geq Z(\omega)$.
\end{enumerate}
\end{theorem}


\subsection{Determination of the zonal enstrophy level}
\label{subsec:estimate_lambda}
To apply Theorem\,\ref{theorem:laminated} to the estimation of attainable zonal enstrophy, 
we have to determine the eigenvalue $\lambda$ that identifies the zonal enstrophy level.
Here, we suggest the following method (which we will examine and improve in Sec.\,\ref{sec:simulation}).

The self-organization of zonal flow can be seen as a relaxation process of the zonal enstrophy level,
which parallels the inverse cascade in the meridional wave number space.
Just as the transition of the quantum energy level is caused by photon emission, the relaxation of the zonal enstrophy level is due to the emission of wavy vorticity, which is driven by the nonlinear coupling of the zonal and wavy components.
Therefore, the relaxation 
can proceed as far as the nonlinear term $\{\omega, \psi\}$ dominates the evolution equation (\ref{beta-eq}).
Relative to the concomitant linear term $\beta\{y,\psi\}$,
the nonlinear term becomes weaker as the length scale increases (i.e., the inverse cascade proceeds).
On the Rhines scale\,\citep{Rhine}
\begin{equation}
L_R = \sqrt{\frac{2U}{\beta}}, 
\label{Rhines-original}
\end{equation}
the linear and nonlinear terms have comparable magnitudes,
where $U$ is the representative magnitude of the zonal flow velocity. 

Since the energy is conserved, we may estimate $U = \sqrt{2C_E}$.
Hence, we have an \emph{a priori} estimate
\begin{equation}
\lambda \sim \frac{\pi}{L_R}= \pi\sqrt{\frac{\beta}{2\sqrt{2C_E}}}.
\label{Rhines-original-2}
\end{equation}
Notice the influence of the energy $C_E$ on the eigenvalue $\lambda$.
Although each value of the zonal enstrophy level is independent to $C_E$, the selection of the level is made by $C_E$.

In the next section, we will examine the theoretical estimates by comparing numerical simulation results.

\section{Comparison with numerical simulations}\label{sec:simulation}
\subsection{Simulation model}
In this section, we compare the forgoing theoretical estimates with numerical simulation results. 
With a system size $L$ and a rotation period $T$,
we normalize the variables as
\begin{equation}
\check{x} = \frac{x}{L}, \ \check{y} = \frac{y}{L}, \ \check{t} = \frac{t}{T}, \ \check{\omega}=\omega T, \ \check{\psi} \ = \frac{\psi T}{L^2},
\end{equation}
by which the vorticity equation reads
\begin{equation}
\partial_{\check{t}} \check{\omega} + \{\check{\omega} + \beta \check{y}, \check{\psi} \} = \nu \nabla \check{\omega},
\end{equation}
where $\nu$ represents the viscosity (reciprocal Reynolds number).
For simplicity, we will omit the normalization symbol $\check{~}$ in the following description. 
Whereas our theoretical analysis is based on the dissipation-free model (\ref{beta-eq}),
we add a finite viscosity $\nu$ for numerical stability
(typically, we put $\nu=1.0\times10^{-6}$).
A finite viscosity is also indispensable for the self-organization process, because the ideal (zero viscosity) dynamics is constrained by infinite number of Casimirs (local circulations), preventing changes in streamline topology.
The theoretical model, however, ignores the dissipation by assuming the robustness of the invariants that are used as constraints (see Proposition\,\ref{proposition:constants}). 
The influence of dissipation will be examined carefully when we compare the theory and numerical simulation.

In the following simulation, we assume parameters comparable to the Jovian atmosphere, where
$L=4.4 \times 10^8$m, $T=8.6 \times 10^5$sec.
The parameter $\beta$ is determined as
\[
\beta=\frac{2\Omega}{R}(\cos \theta) LT,
\]
where $\Omega$ is the angular vorticity of rotating frame and $R$ is the radius and $\theta$ is latitude.
For $L\sim 2\pi R $ and $\theta \sim 0$, we obtain $\beta \sim 10^2$. 
The jet velocity reaches $U\sim1\times10^2$ m/s, which yields $C_F\sim4\times10^{-1}$ and $C_L\sim2\times10^{-1}$ if the jet achieves the maximum opposite velocities on both north and south boundaries. 
Here we assume moderate values $C_F\sim10^{-1}$ and $C_L\sim10^{-1}$.

\subsection{Self-organized zonal flow}\label{sim2}
As we have seen, the theoretical estimate of the minimum $Z(\omega)$ changes dramatically depending on whether $C_F$ and $C_L$ are finite or not (Sec.\,\ref{subsec:W-E-L-F}).
First, we study the general case where both $C_F$ and $C_L$ are finite 
(the special case of $C_F=0$ and $C_L=0$ will be examined in Sec.\,\ref{subsec:degenerate}).
We assume an initial condition such that
\[
\omega|_{t=0} =5.0 \sin 15\pi y + \sum_{m,n} \alpha_{mn} e^{imx} \sin n\pi y ,
\] 
with random $\alpha_{mn} (|\alpha_{mn}| \in [0,50)$ for $5 \le m,n \le 10$),
which yields $C_E=3.6\times10^{-2},\,C_F=0.21$ and $C_L=0.105$.

\begin{figure} [H]
\begin{center}
\includegraphics[scale=0.3]{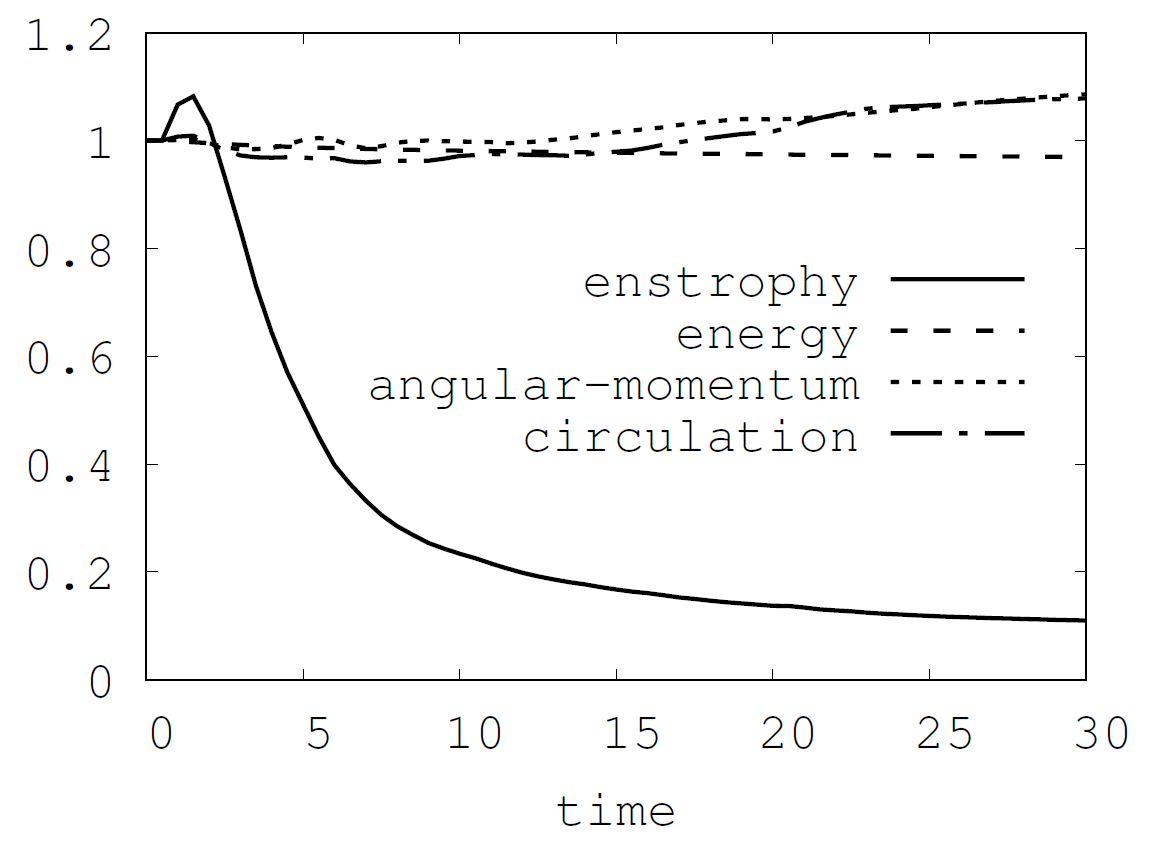}
\end{center}
\caption{
\label{fig:graph4}
The evolution of the ``ideal'' constants in the simulation.
Each value is normalized by the corresponding initial value.
}
\end{figure}

In Fig.\,\ref{fig:graph4}, we show the evolution of the ``ideal'' constants.
The total energy $C_E$ is well conserved.
The changes in $C_F$ and $C_L$ are also tolerable. 
Because of a finite viscosity ($\nu=1.0\times10^{-6}$), however, the total enstrophy $C_Q$ changes significantly.
But it is not essential for the present purpose of comparison,
because the theoretical estimate of minimum $Z(\omega)$ is independent of the $C_Q$.
As noted after (\ref{k_lambda})-(\ref{B_lambda}),
the total enstrophy $Q(\omega)=C_Q$ only contributes to estimating the wave number $k$ of the wavy component $\omega_w$.
As the simulation shows, the ``dissipation'' of the total enstrophy is the signature of the relaxation,
when we consider a finite viscosity. 
We may interpret the dissipation as the scale separation between the visible scale and micro scale;
the latter is separated from the vortex dynamics model by suppressing the amplitudes of micro-scale vortices.
This scenario is consistent with the local interaction model;
the nonlinear dynamics is dominated by interactions among similarly sized vortices (i.e., local in the Fourier space) within the inertial range, so it is not influenced by vortices of far smaller scales.



\begin{figure} [H]
\begin{center}
\includegraphics[scale=0.6]{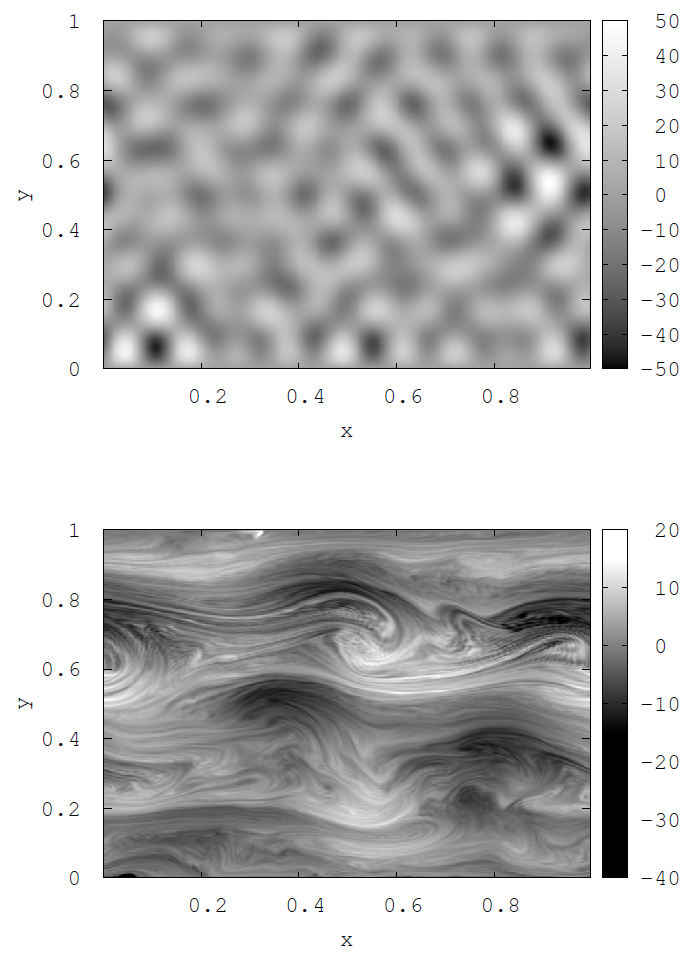}
\end{center}
\caption{
\label{fig:graph5}
Self-organization of zonal flow (gray level represents to the local value of $\omega$).  
(left) Initial condition with finite circulation $C_F=0.21$ and angular momentum $C_L=0.11$.
(right) Creation of zonal flow observed at $t=20$.
}
\end{figure}

\begin{figure} [H]
\begin{center}
\includegraphics[scale=0.5]{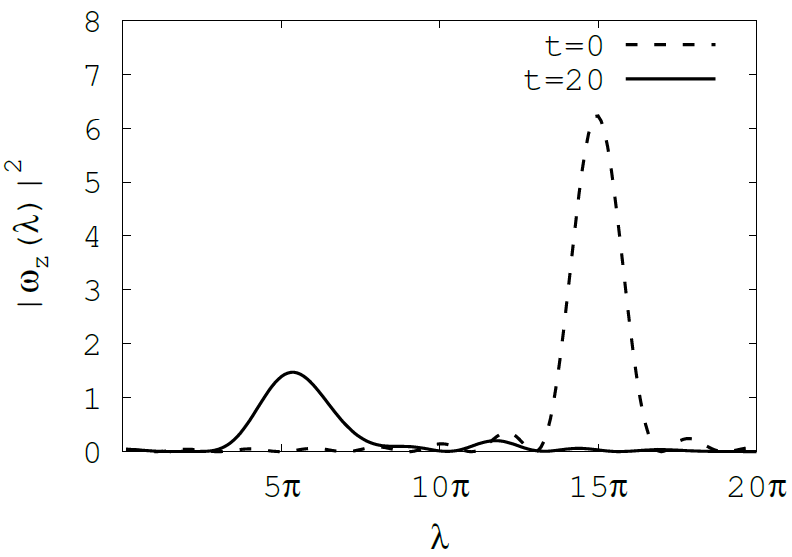}
\end{center}
\caption{
\label{fig:graph6}
The Fourier spectrum of the zonal vorticity $\omega_z=\mathcal{P}_z \omega$ in the self-organized state ($t=20$).
The eigenvalue $\lambda \sim 5\pi$ is dominant.
}
\end{figure}

Fig.\,\ref{fig:graph5} shows the self-organized state ($t=20$), where an appreciable zonal component manifests.
In Fig.\,\ref{fig:graph6}, we compare the Fourier spectrum of the zonal component $\omega_z = \mathcal{P}_z\omega$ in the initial and self-organized states.
We find the redistribution of the spectrum into lower $\lambda$ modes (i.e., inverse cascade).
A comparison with the Rhines scale will be described later.


\begin{figure} [H]
\begin{center}
\includegraphics[scale=0.35]{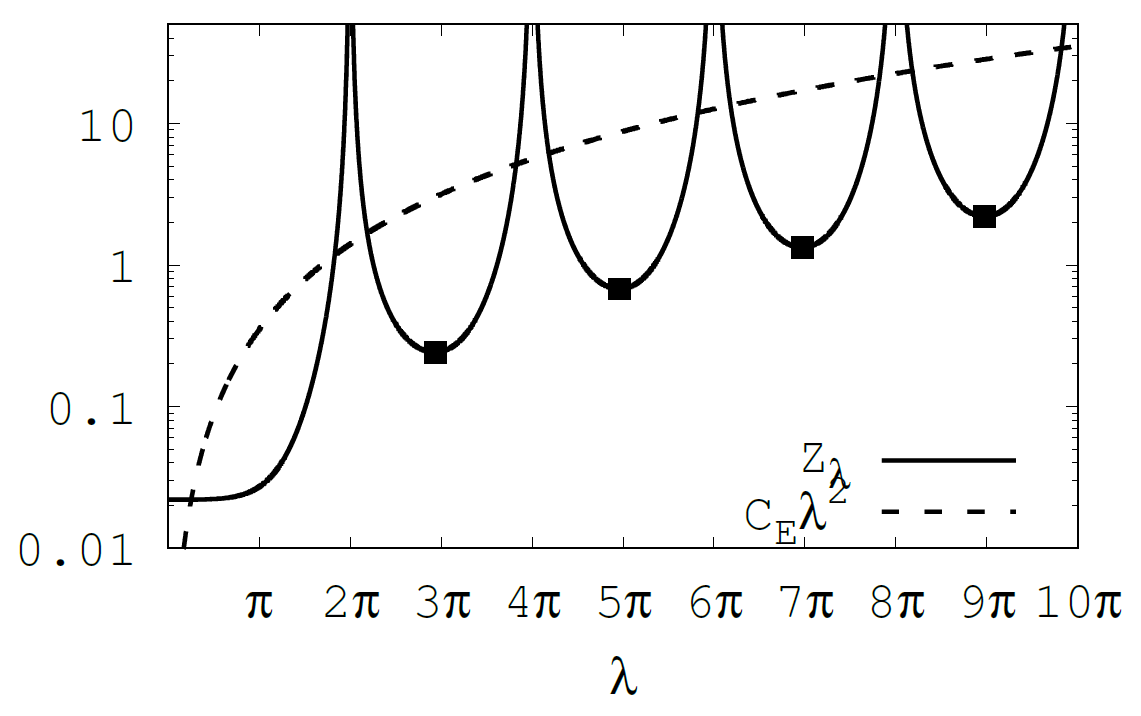}
\end{center}
\caption{
\label{fig:graph7}
The graphs of $Z_\lambda$ and $C_E \lambda^2$, 
evaluated for the parameters corresponding to the simulation of Fig.\,\ref{fig:graph5}. 
The points on the curve indicates the eigenvalues.
}
\end{figure}

\begin{figure} [H]
\begin{center}
\includegraphics[scale=0.45]{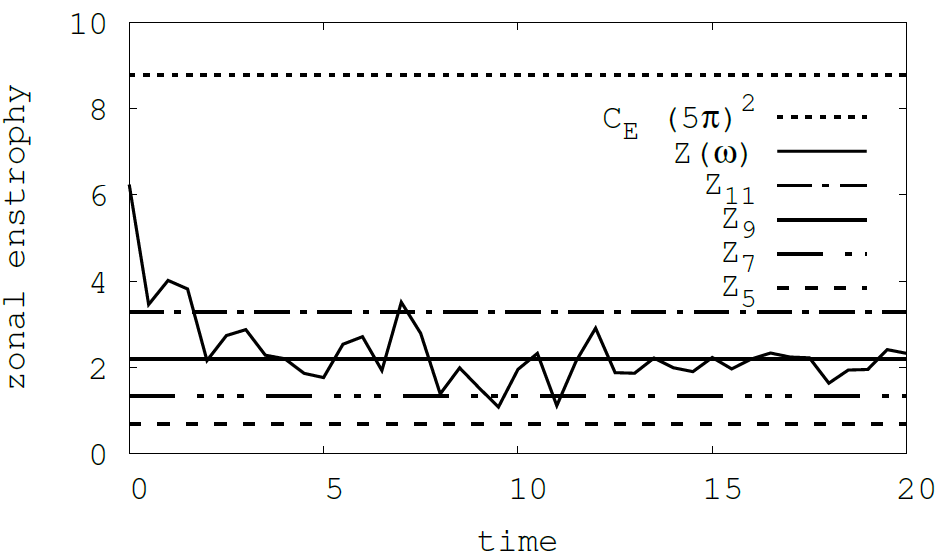}
\end{center}
\caption{
\label{fig:graph8}
Evolution of the zonal enstrophy $Z(\omega)$, and its comparison with the theoretical minimum $Z_\lambda$ and the maximum $C_E \lambda^2$ evaluated for the self-organized state $\lambda \sim 5\pi$. 
To demonstrate the sensitivity of the minimum value, we also show the theoretical minimum $Z_\lambda$ evaluated for $\lambda \sim 5\pi$, $7\pi$, $9\pi$ and $11\pi$. 
}
\end{figure}

To make comparison with the theoretical estimate of zonal enstrophy, we plot
$Z_\lambda$ (the theoretical minimum of zonal enstrophy) and 
$C_E \lambda^2$ (the theoretical maximum of zonal enstrophy), 
evaluated for the parameters determined by the given initial condition,
in Fig.\,\ref{fig:graph7}.
As $\lambda=5\pi$ is the dominant mode (Fig.\,\ref{fig:graph6}), we obtain
$Z_\lambda = 0.69$ and $C_E \lambda^2 = 8.8$.
In Fig.\,\ref{fig:graph8}, we compare the simulation result and the theoretical estimates,
demonstrating that the actual zonal enstrophy $Z(\omega)$ stays between the theoretical minimum and maximum;
the estimate of the lower bound is reasonably accurate.


\subsection{Improved Rhines scale}
\label{sec:Rhines-improved}
\begin{figure} [H]
\begin{center}
\includegraphics[scale=0.375]{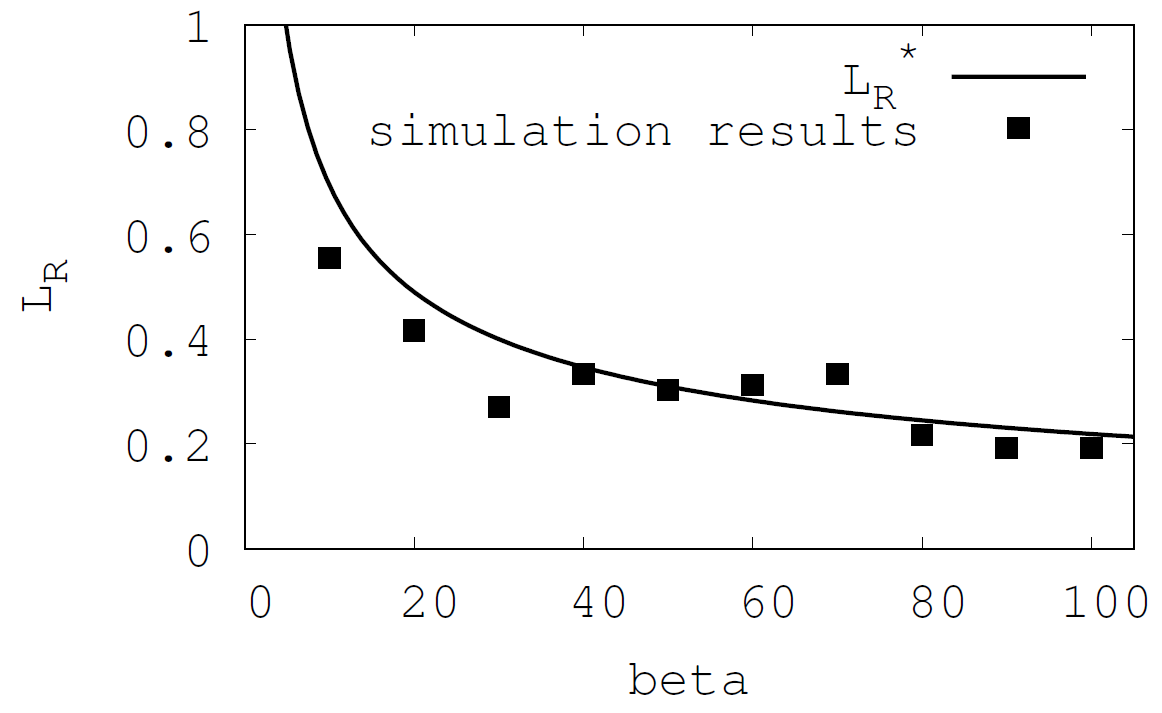}
\end{center}
\caption{
\label{fig:graph9}
The comparison between the dominant scale in the self-organized state and the Rhines scale for different values of $\beta$. $L^*_R$ denotes the modified Rhines scale.
}
\end{figure}

The forgoing discussion depends on the \emph{a posteriori} estimate of the eigenvalue $\lambda$.
As discussed in Sec.\,\ref{subsec:estimate_lambda},
however, we need an \emph{a priori} estimate of $\lambda$ to make the theory useful.
While the Rhines scale $L_R$ has been proposed to estimate $\lambda \sim \pi/L_R$,
it turns out to be too crude.
Here, we propose an improved Rhines scale to make more accurate estimate. 
Figure\,\ref{fig:graph9} compares the dominant scale in the final state obtained by simulation and the Rhines scale for different values of $\beta$. 
It is shown that the dominant scale is approximately 3 times of the Rhines scale.

The Rhines scale (\ref{Rhines-original}) is the length scale $L_R$ at which the magnitudes of the nonlinear term $\{\omega,\psi\}$ and the linear term $\beta\{ y, \psi \}$ become comparable.
However, it seems that the function of the nonlinear term, that derives the relaxation of the enstrophy level $\lambda$,
does not end immediately at $L_R$;
the numerical experiment shows that the relaxation continues up to $\sim 3\times L_R$,
where the magnitude of the nonlinear term becomes about one eighth of the linear term.
Therefore, we propose to use $L^*_R = 3 L_R$ for the \emph{a priori} estimate $\lambda = \pi/L^*_R$;
modifying (\ref{Rhines-original-2}), we estimate
\begin{equation}
\lambda \sim \frac{\pi}{3} \sqrt{\frac{\beta}{2\sqrt{2C_E}}}.
\label{Rhines-improved-2}
\end{equation}


\subsection{Degenerate case: $C_F=0$ and $C_L=0$}
\label{subsec:degenerate}
Finally, we examine the degenerate case of $C_F=0$ and $C_L=0$,
where we cannot provide nontrivial estimate of the minimum zonal enstrophy
(see Theorem\,\ref{theorem:laminated}-2).
However, we still observe self-organization of zonal flow, and the corresponding enstrophy satisfies the maximum condition.

\begin{figure} [H]
\begin{center}
\includegraphics[scale=0.6]{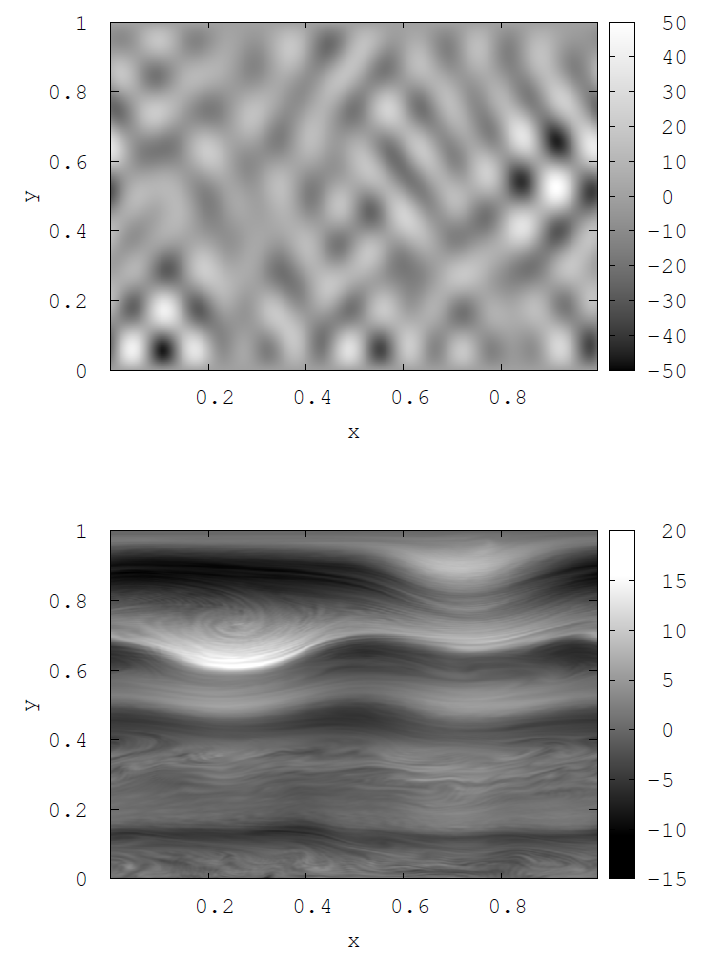}
\end{center}
\caption{
\label{fig:graph10}
Self-organization of zonal flow (gray level corresponds to the local value of $\omega$).  
(left) Initial condition with zero zonal component $\omega_z=0$.
(right) Creation of zonal flow observed at $t=50$. 
}
\end{figure}

Figure\,\ref{fig:graph10} shows the creation of zonal flow from an initial condition
\[
\omega|_{t=0} =\sum_{m,n} \alpha_{mn} e^{imx} \sin n\pi y,
\] 
with random $\alpha_{mn} (|\alpha_{mn}| \in [0,50)$ for $5 \le m,n \le 10$)
which is free from zonal component ($\omega_z=0$ at $t=0$).
The symmetry also yields $C_F=0$ and $C_L=0$,
so that the special condition of Theorem\,\ref{theorem:laminated}-2 applies.
We only have a nontrivial estimate of the upper bound of $Z(\omega)$.

\begin{figure} [H]
\begin{center}
\includegraphics[scale=0.35]{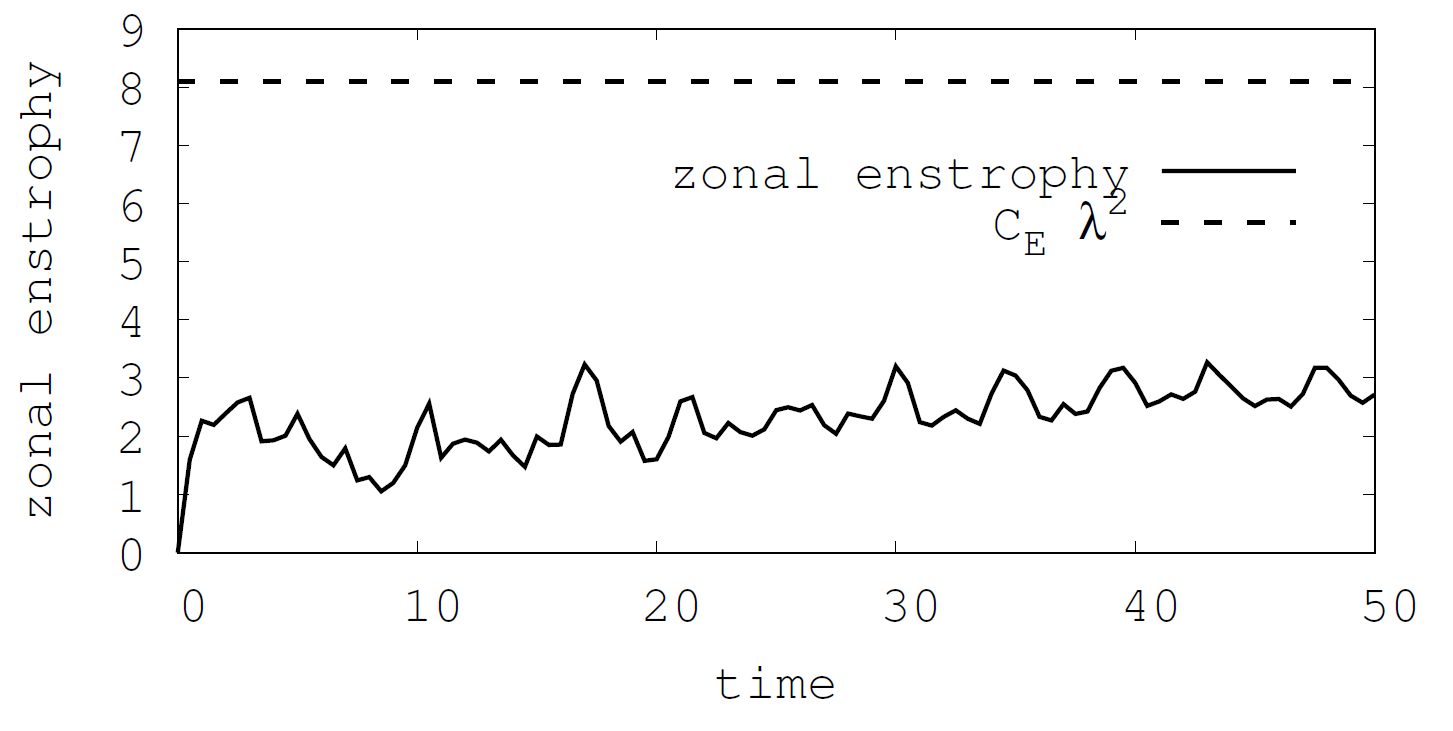}
\end{center}
\caption{
\label{fig:graph11}
Evolution of the zonal enstrophy $Z(\omega)$ and its comparison with the theoretical estimate (upper bound).
}
\end{figure}

\begin{figure} [H]
\begin{center}
\includegraphics[scale=0.3]{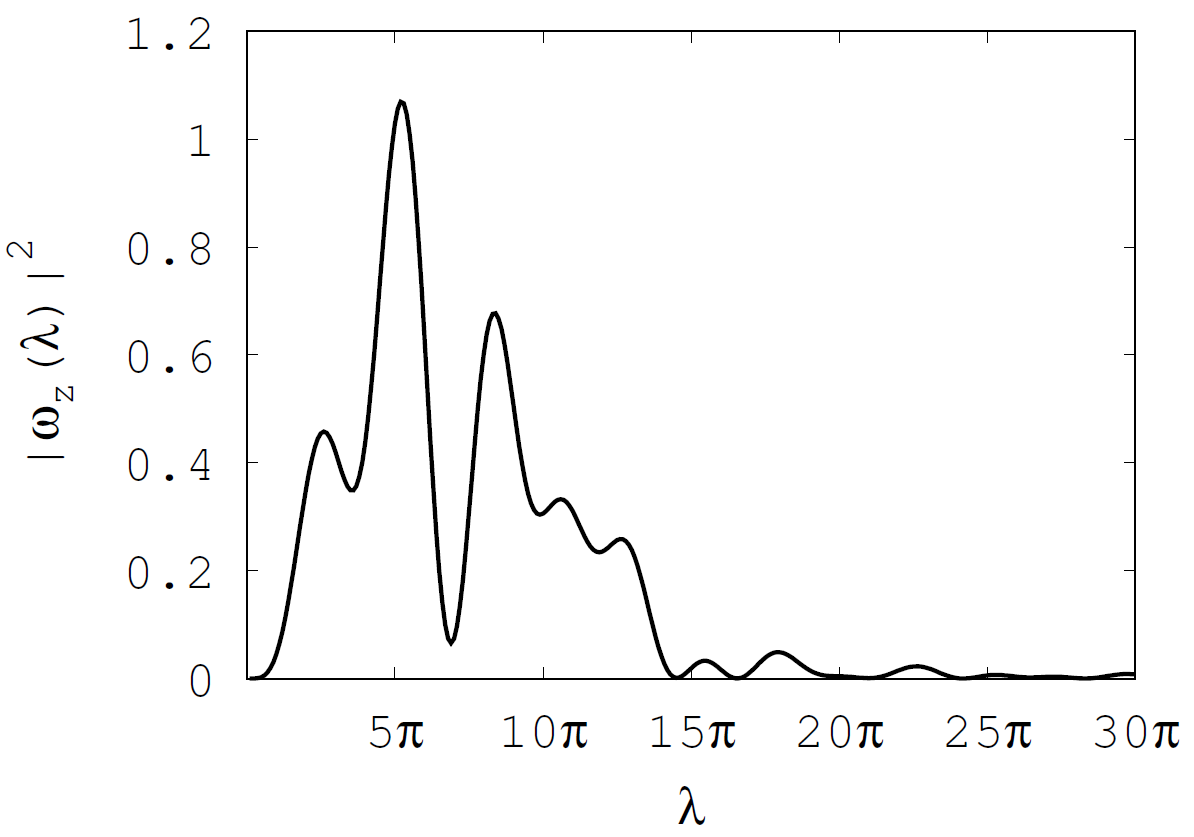}
\end{center}
\caption{
\label{fig:graph12}
Fourier spectrum of the zonal vorticity in the self-organized state (Fig.\,\ref{fig:graph10} (right)).
}
\end{figure}

In Fig.\,\ref{fig:graph11}, we plot the evolution of the zonal enstrophy $Z(\omega)$,
and compare it with the theoretical maximum (\ref{Z-max-2}).
Here we used $\lambda = 5\pi \sim 1/L^*_R$ by the improved Rhines estimate;
in Fig.\,\ref{fig:graph12}, we show the Fourier spectrum of $\omega_z$,
which supports the choice.
We observe that the time-asymptotic value of $Z(\omega)$ stays below the upper bound $C_E \lambda^2$.


\section{Conclusion}
\label{sec:conclusion} 
\begin{figure} [H]
\begin{center}
\includegraphics[scale=0.3]{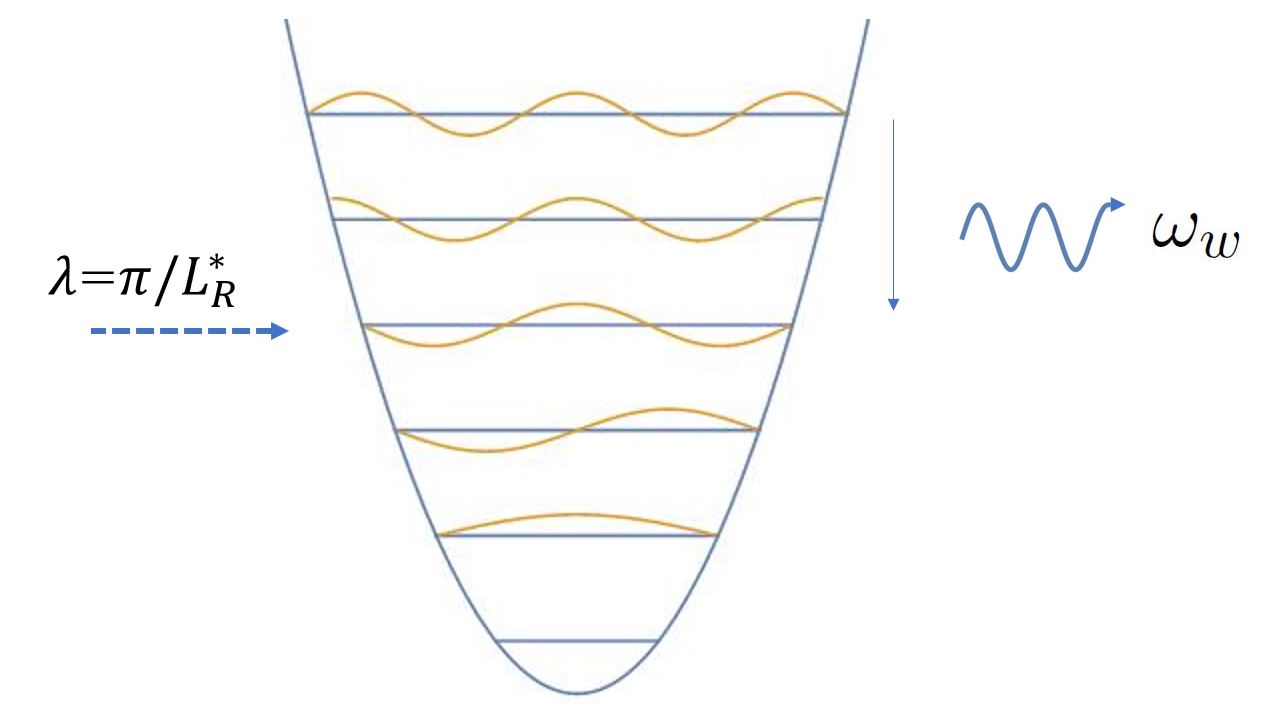}
\end{center}
\caption{
\label{fig:analogy}
Analogy of quantum energy levels and ``deexcitation'' by emitting small-scale wavy enstrophy, which parallels the forward cascade of enstrophy.
}
\end{figure}

We have found a discrete set of \emph{zonal enstrophy levels}
that are quantized by the eigenvalue $\lambda$ measuring the lamination period (= system size in the latitude direction / lamination number).
As shown in Fig.\ref{fig:graph3}, a finite circulation $C_F$ and/or angular momentum $C_L$ bring about symmetry breaking in the eigenstates (minimizers), inhibiting even number laminations.
In actual situation, however, the \emph{mixed state} may include spectra of even number laminations (see Figs.\,\ref{fig:graph6} and \ref{fig:graph12}).
By comparing with simulation results, we verified that the theoretical value $Z_\lambda$ gives a reasonable estimate of the zonal enstrophy, if we choose the relevant lamination number.
We note that the enstrophy levels are determined independently of the selection mechanism (see Appendix B).
Just as the quantum energy level of an orbital electron is lowered by photon emission (see Fig.\,\ref{fig:analogy}), the relaxation of the zonal enstrophy level proceeds by the emission of short-scale wavy vorticity.
The relaxation process can be viewed as the forward cascade of enstrophy (creation of short-scale wavy vortices) and the simultaneous inverse cascade of the energy spectrum (deexcitation to lower zonal enstrophy states).
The energy constraint plays an essential role in selecting the level;
the relaxation continues as far as the nonlinear term, measured by the energy, dominates the evolution. 
The Rhines scale estimates the balance point of the magnitudes of the nonlinear term and the linear Rossby wave term,
but we found that the nonlinear effect continues to work until it becomes about one order of magnitude smaller than the linear term, so we propose an improved Rhines scale.

Comparing Theorems\,\ref{prop:linear} and \ref{theorem:laminated}, 
we find that the energy constraint $E(\omega)=C_E$ plays an essential role in creating the discrete zonal enstrophy levels  $Z_\lambda$.
Interestingly, the value $C_E$ does not influence the value of each zonal enstrophy $Z_\lambda$,
which is determined only by the other constants $C_F$ (circulation) and $C_L$ (angular momentum).
However, in absence of the energy constraint, we only have the ``ground state'' $\lambda=0$ as given in Theorem\,\ref{prop:linear}.
In the eigenstate of $\omega_z$ (belonging to the eigenvalue $\lambda$),
the zonal enstrophy $Z(\omega_z)$ and the zonal energy $E(\omega_z)$ are related by $Z(\omega_z)=\lambda^2 E(\omega_z)$. 
Under the energy constraint (and a fixed $\lambda$), therefore,
$Z(\omega_z)$ may take a smaller value when the wavy component $\omega_w$ shares a larger energy $E(\omega_w)$.
The simultaneous total enstrophy constraint contributes to determining the wave number $k$ of the wavy component $\omega_w$.
Without the symmetry breaking constraints by the circulation $F(\omega)$ and the angular momentum $L(\omega)$,
$E(\omega_z)$ can minimize to zero (see Fig.\,\ref{fig:graph1}), and then, $k\rightarrow\infty$ (independently of the specific value of $C_Q$).
Finite symmetry breaking by $C_F$ and/or $C_L$ brings about a non-trivial minimum $E(\omega_z)$ (and the corresponding $Z(\omega_z)=\lambda^2 E(\omega_z)$).  
Then, the partition of the energy to the wavy component is determined as $E(\omega_w)=C_E- E(\omega_z)$,
and the wave number $k$ is determined by $k^2 E(\omega_w) = C_Q-Z(\omega_z)$.
Interestingly, we may not remove the total enstrophy constraint from the variational principle, in order to retain a finite wavy component as the complementary to the zonal component, while its role is limited to characterizing only the wavy component.
This unusual phenomenon in the variational principle is caused by the non-coerciveness of the target functional $Z(\omega)$ with respect to the norm $\|\omega\|$.  

\section*{Acknowledgment}
The authors thank Yohei Kawazura for his support in the simulation study.
This work was supported by JSPS KAKENHI grant number 17H01177.

\appendix
\section{The ABC of variational principle}\label{appendix:variational_principle}
To see the mathematical non-triviality of the variational principle for the zonal enstrophy, we review the standard relation between the target functional and constraint.

\subsection{Target functional and constraint}\label{subsec:TF_Constraint}
We start with a textbook example.
The isoperimetric problem is to (1) maximize the surface area $S$ with a constraint on the periphery length $L$,
or (2) minimize the  peripheral length $L$ with a constraint on the surface area $S$.
Both problems have the same solution, i.e. a circular disk or its periphery.
Notice that reversing the target and constraint in each setting results in an \emph{ill-posed} problem;
one can make $L$ infinitely long without changing $S$, or
one can make $S$ infinitely small without changing $L$.
Let us concentrate on minimization problems.
For a variational principle to be \emph{well-posed}, the target ($L$) must be ``fragile'' and the constraint ($S$) must be robust.
Here the fragility speaks of the sensitivity to small-scale perturbations.
Suppose that we make pleats on a periphery; then $L$ is increased, but $S$ is not necessarily changed.
In analytical formalism, a fragile functional includes a larger number of differentiations
---derivatives are sensitive to small-scale perturbations.
In the forgoing example, we may formally write
\[
S = \int_{\mathbb{R}^2} \mathbb{I}_M \mathrm{d}^2 x,
\quad 
L= \int_{\mathbb{R}^2} |\bnabla \mathbb{I}_M| \mathrm{d}^2 x,
\quad
\mathbb{I}_M(\boldsymbol{x}) = \left\{ \begin{array}{ll}
1 & \mathrm{if}~\boldsymbol{x} \in M
\\
0 & \mathrm{if}~\boldsymbol{x} \notin M,
\end{array} \right.
\]
where $M$ is a simply-connected domain $\subset\mathbb{R}^2$ that should be optimized to minimize $L$ for some given value of $S$.
Including $\bnabla$ in the integrand, $L$ is fragile.

\subsection{Coerciveness and continuity}\label{subsec:coercivity}
To make the argument more precise, we introduce the notion of \emph{coercive functionals};
cf.\,\citep{VPtextbook,YM}.
Let $u$ be a real-valued function (state vector) belonging to a function space (phase space) $V$,
which is a Banach space with a norm $\|u \|$.
A real-valued functional $G(u)$ is said coercive, if 
\begin{equation}
\| u \|^2 \leq c\, G(u),
\label{coercive}
\end{equation}
where $c$ is some positive constant.
On the other hand, a real-valued functional $H(u)$ is continuous, if
\begin{equation}
| H(u+\delta) - H(u) | \rightarrow 0 \quad (\| \delta \| \rightarrow 0).
\label{continuous}
\end{equation}
We can formulate a well-posed minimization problem with a coercive target functional $G(u)$ and a continuous constraining functional $H(u)$ 
(we may also consider multiple constraints with continuous functionals).

To see how the coerciveness and continuity influence variational principles, let us consider an example with two functionals
\[
{G}(u) = \int_M |\bnabla u(\boldsymbol{x})|^2 \mathrm{d}^n x,
\quad
{H}(u) = \int_M |u(\boldsymbol{x})|^2 \mathrm{d}^n x ,
\]
where $u$ is a scalar function defined in a smoothly bounded open set $M\subset\mathbb{R}^n$.
We assume that $u=0$ on the boundary $\partial M$.
Notice that ${H}(u)^{1/2}$ is the $L^2$ norm $\| u \|$.
Therefore, $H(u)$ is a continuous functional on the function space $V=L^2(M)$.
By the Poincar\'e inequality, we have
\[
\| u \|^2 \leq c \| \bnabla  u \|^2 = c G(u)
\]
with a positive constant $c$.  Therefore, $G(u)$ is a coercive functional.
 
First, we seek for a minimizer of ${G}(u)$
with the constraint ${H}(u) =1$.
This is a well-posed problem.
The minimizer is found by the variational principle
\begin{equation}
\delta [{G}(u) - \lambda{H}(u)] =0,
\label{prototypeVP}
\end{equation}
where $\lambda$ is a Lagrange multiplier.
The Euler-Lagrange equation 
\[
-\Delta u = \lambda u,
\]
together with the above-mentioned boundary condition, constitute an
eigenvalue problem. We can easily show that every eigenvalue $\lambda$ is positive.
Let $\lambda_j$ be an eigenvalue and $\varphi_j$ be the corresponding normalized
eigenfunction ($\| \varphi_j\|^2  =1$).
With setting $u = a\varphi_j$, and demanding ${H}(u)=1$, we obtain $a=1$ and
${G}(u) = \lambda_j$. The smallest $\lambda_j$, then, yields the 
minimum ${G}(u)$.

The reversed problem of finding
a minimizer of ${H}(u)$ with the restriction ${G}(u) =1$ is ill-posed,
because the constraint is posed by a functional $G(u)$ that is not continuous in the topology of $L^2(M)$.
Let us elucidate the pathology.
The variational principle $\delta [{H}(u) - \mu{G}(u)] =0$
($\mu$ is a Lagrange multiplier) yields the 
Euler-Lagrange equation $-\Delta u = \mu^{-1} u$.
Let $\mu^{-1} = \lambda_j$ (an eigenvalue of $-\Delta$), and $u=a \varphi_j$.
The condition ${G}(u) =1$ yields $a=\lambda_j^{-1/2}$, and
${H}(u) = 1/\lambda_j$.  Hence, the minimum of ${H}(u)$ is achieved by 
the largest eigenvalue that is unbounded, viz., $\mathrm{inf} {H}(u) = 0$ and 
the minimizer $\lim_{\lambda_j \rightarrow\infty} \lambda_j^{-1/2} \varphi_j = 0$
is nothing but the minimizer of ${H}(u)$ without any restriction.
The constraint ${G}(u) =1$ plays no role in this minimization problem.

\subsection{Non-coercive target functional}\label{subsec:non-coercive}
Let us modify the target functional of (\ref{prototypeVP}) to a non-coercive functional.
Let $V_k = \mathrm{span}\,\{ \varphi_1, \cdots, \varphi_k \}$, which is a closed (finite-dimension) subspace of $V=L^2(M)$.
We denote the orthogonal complement by $V'$, i.e. we decompose $V= V_k \oplus V'$.
Let $\mathcal{P}$ be the orthogonal projector $V \rightarrow V'$.
Consider
\[
G' (u) = \| \bnabla (\mathcal{P} u)\|^2 = \int_M (-\Delta \mathcal{P} u) (\mathcal{P} u)\,\mathrm{d}^n x =
\sum_{j>k}^\infty \lambda_j (u, \varphi_j)^2 ,
\]
where $(f,g) = \int_M f(\boldsymbol{x})\,g(\boldsymbol{x}) \,\mathrm{d}^n x$ is the inner product of $L^2(M)$.
Evidently, $G'(u)$ is not coercive.
The modified variational principle
\begin{equation}
\delta [{G}' (u) - \lambda{H}(u)] =0
\label{prototypeVP-truncated}
\end{equation}
yields the Euler-Lagrange equation that reads, after expanding with eigenfunctions,
\begin{equation}
\lambda'_j (u,\varphi_j) = \lambda (u,\varphi_j)
\quad (j=1,2,\cdots),
\label{EL-truncated}
\end{equation}
where the ``modified eigenvalues'' are 
\[
\lambda'_j = \left\{ \begin{array}{ll}
 0 &  (j=1,\cdots, k),
\\
 \lambda_j &  (j> k).
\end{array} \right.
\]
The minimizer of $G'(u)$ is a solution of (\ref{EL-truncated}) such that 
$\lambda=0$ and 
\[
u = \sum_{j=1}^k a_j \varphi_j,
\]
where constants $a_1,\cdots,a_k$ can be arbitrarily chosen provided  that
$ \sum_{j=1}^k |a_j|^2 = 1$ in order to satisfy the constraint $H(u)=1$.
We obtain $\mathrm{min}\,G' = 0$, but the minimizer is not a unique function.

This prototypical example elucidates the essence of the pathology created in a variational principle with non-coercive target functional.
We encounter a similar non-uniqueness (degeneracy) problem in Sec.\,\ref{subsec:W}.
Interestingly, however, the energy constraint brings about a dramatical change in the mathematical structure, and removes the degeneracy; Sec.\,\ref{subsec:W-E}.

\section{Comparison with the minimization of total enstrophy}
\label{appendix:selective-dissipation}
\begin{figure} [H]
      \begin{center}
      \includegraphics[scale=0.4]{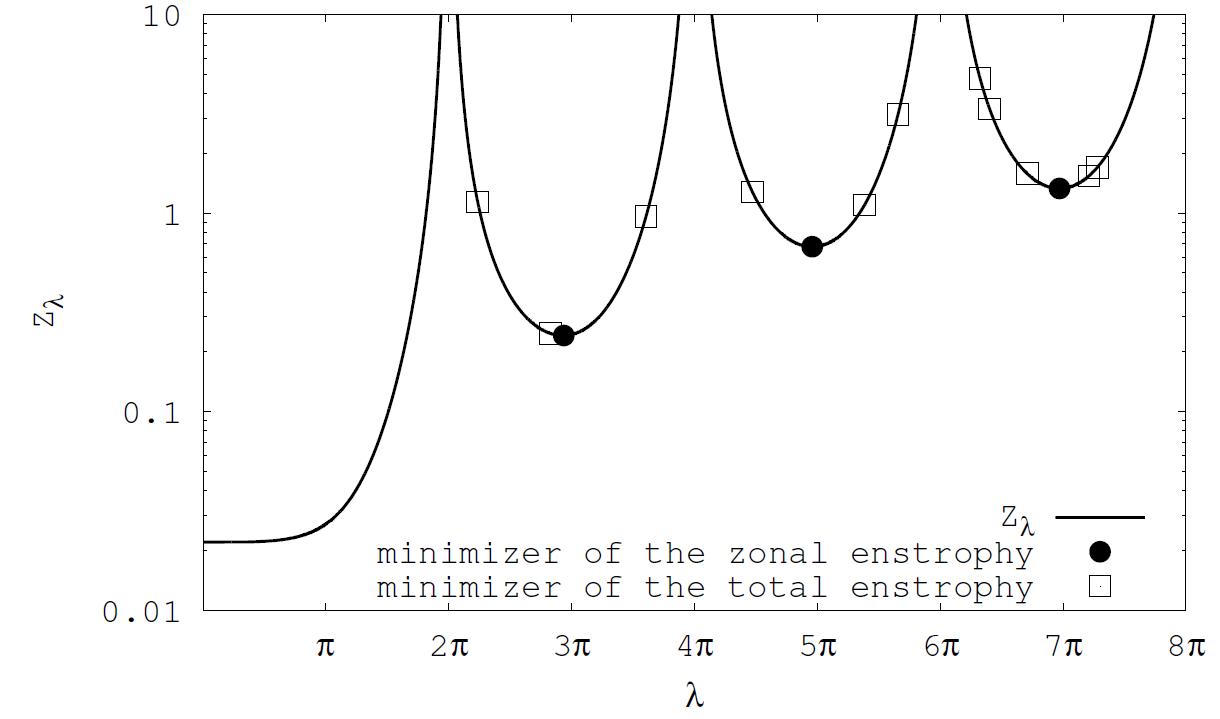}
      \end{center}
      \caption{
      \label{fig:graph5-6-1}
      Comparison of the minimum zonal enstrophy (circle points; as given in Fig.\ref{fig:graph3})
      and the zonal enstrophy $Z(\omega)$ evaluated for the minimizer of the total enstrophy (given by $\psi_z$ of (\ref{psi_z-6}) (square points).
      The parameters correspond to those used for the simulation of Fig.\,\ref{fig:graph5}. 
      }
\end{figure}

While the physical meaning of the variational principle for the zonal enstrophy is very different from that of the total enstrophy, it may be of mathematical interest to compare their solutions
(the latter has been studied in the context of  \emph{selective dissipation}\,\citep{BH76})
Let us seek the minimizer of 
\[
Q(\omega) - \mu_0 F(\omega) -\mu_1 L(\omega) - \mu_2 E(\omega).
\]
The Euler-Lagrange equation is
\begin{equation}
\omega -\mu_0 -\mu_1 y - \mu_2 \mathcal{K}\omega = 0.
\end{equation}
Imposing the boundary conditions, 
we obtain $\psi=\psi_z + \psi_w$ with
\begin{eqnarray}
\psi_z &=& A_1 \cos \lambda y + A_2 \sin \lambda y - \frac{\mu_0 + \mu_1 y}{\mu_2} ,\label{psi_z-6}\\
\psi_w &=& B \sin k_x x \ \sin k_y y\label{psi_w-6},
\end{eqnarray}
where
\begin{equation}
      \lambda^2=k_x^2+k_y^2=\mu_2,\label{lambda-k}
\end{equation}
and 
\begin{equation}
      k_x = 2 n_2 \pi , \ k_y = n_3 \pi \ (n_2,\ n_3 \in \mathbb{Z}).\label{k-n}
\end{equation}
Notice that both $\psi_z$ and $\psi_w$ are similar to those of the zonal enstrophy minimizer given in (\ref{zonal-potential2})-(\ref{wave-potential2}).
However, the eigenvalue $\lambda$ is determined differently.
Here, (\ref{lambda-k}) directly gives $\lambda$, 
while the eigenvalues for the zonal enstrophy is determined as the local minimums of the graph of $Z_\lambda$;
see Fig.\ref{fig:graph3}.
In Fig.\,\ref{fig:graph5-6-1}, we compare these two different results;
circle points, repeating those of Fig.\ref{fig:graph3}, show the minimum zonal enstrophy,
while square points show $Z(\omega)$ evaluated for $\psi_z$ of (\ref{psi_z-6}).
We see how the minimum zonal enstrophy is always the \emph{minimum}.
If the selective dissipation applies (see Sec\,\ref{sec:introduction}), the relaxed state arrives at one of the red points.
Other mechanisms omitted in the selective dissipation model may diminish the zonal enstrophy further, but it cannot go below the minimum zonal enstrophy. 


\bibliographystyle{jfm}
\bibliography{enstrophy}

\end{document}